\documentclass[onecolumn,draftclsnofoot]{IEEEtran}
\usepackage{amsmath}
\usepackage{graphicx}
\usepackage{caption2}
\usepackage{amsthm}
\usepackage{float}
\usepackage{mathrsfs}
\usepackage{verbatim}
\usepackage{epstopdf}
\usepackage{amssymb}
\usepackage{amsfonts}
\usepackage{subfigure}
\usepackage{color}
\usepackage{cancel}
\usepackage{amsmath,amssymb,amsthm,mathrsfs,amsfonts,dsfont}
\usepackage[shortlabels]{enumitem}
\DeclareMathOperator*{\argmin}{arg\,min}

\newcommand\blfootnote[1]{%
  \begingroup
  \renewcommand\thefootnote{}\footnote{#1}%
  \addtocounter{footnote}{-1}%
  \endgroup
}

\begin{document}
\title{Gaussian Multiple Access Channels with One-Bit Quantizer at the Receiver}
\author{{Borzoo Rassouli, Morteza Varasteh and Deniz G\"{u}nd\"{u}z} %
\thanks{Borzoo Rassouli, Morteza Varasteh and Deniz G\"{u}nd\"{u}z are with the Intelligent Systems and Networks group of Department of Electrical and Electronics,
Imperial College London, United Kingdom. emails: \{b.rassouli12, m.varasteh12, d.gunduz\}@imperial.ac.uk.}}
\maketitle
\begin{abstract}
%\boldmath
The capacity region of a two-transmitter Gaussian multiple access channel (MAC) under average input power constraints is studied, when the receiver employs a zero-threshold one-bit analog-to-digital converter (ADC). It is proved that the input distributions of the two transmitters that achieve the boundary points of the capacity region are discrete. Based on the position of a boundary point, upper bounds on the number of the mass points of the corresponding distributions are derived.
%Furthermore, a closed-form solution for the sum capacity of this channel is derived, and it is shown that time division with power control achieves it. Finally, a conjecture on the sufficiency of $K$ mass points in a point-to-point real AWGN with a $K$-bin ADC front end at the receiver (symmetric or asymmetric) is also settled.
\end{abstract}
\begin{IEEEkeywords}
Gaussian multiple access channel, one-bit quantizer, capacity region\blfootnote{This work has been presented partially in \cite{ISIT_17}.}.
\end{IEEEkeywords}
\section{Introduction}
The energy consumption of an analog-to-digital converter (ADC) (measured in Joules/sample) grows exponentially with its resolution (in bits/sample) \cite{Walden_1999}, \cite{Murmann_2014}. When the available power is limited, for example, for mobile devices with limited battery capacity, or for wireless receivers that operate on limited energy harvested from ambient sources \cite{deniz}, the receiver circuitry may be constrained to operate with low resolution ADCs. The presence of a low-resolution ADC, in particular a one-bit ADC at the receiver, alters the channel characteristics significantly. Such a constraint not only limits the fundamental bounds on the achievable rate, but it also changes the nature of the communication and modulation schemes approaching these bounds. For example, in a real additive white Gaussian noise (AWGN) channel under an average power constraint on the input, if the receiver is equipped with a $K$-bin (i.e., $\log_2K$-bit) ADC front end, it is shown in \cite{jaspreet} that the capacity-achieving input distribution is discrete with at most $K+1$ mass points. This is in contrast with the optimality of the Gaussian input distribution when the receiver has infinite resolution.

Especially with the adoption of massive multiple-input multiple-output (MIMO) receivers and the millimeter wave (mmWave) technology enabling communication over large bandwidths, communication systems with limited-resolution receiver front ends are becoming of practical importance. Accordingly, there have been a growing research interest in understanding both the fundamental information theoretic limits and the design of practical communication protocols for systems with finite-resolution ADC front ends. In \cite{Krone_Fettweis_2010}, the authors show that for a Rayleigh fading channel with a one-bit ADC and perfect channel state information at the receiver (CSIR), quadrature phase shift keying (QPSK) modulation is capacity-achieving. In case of no CSIR, \cite{Mezghani_2008} shows that (QPSK) modulation is optimal when the signal-to-noise (SNR) ratio is above a certain threshold, which depends on the coherence time of the channel, while for SNRs below this threshold, on-off QPSK achieves the capacity.
For the point-to-point multiple-input multiple-output (MIMO) channel with a one-bit ADC front end at each receive antenna and perfect CSIR, \cite{Mezghani_2007} shows that QPSK is optimal at very low SNRs, while with perfect channel state information at the transmitter  (CSIT), upper and lower bounds on the capacity
are provided in \cite{Heath}.

To the best of our knowledge, the existing literature on communications with low-resolution ADCs focus exclusively on point-to-point systems. Our goal in this paper is to understand the impact of low-resolution ADCs on the capacity region of a multiple access channel (MAC). In particular, we consider a two-transmitter Gaussian MAC with a one-bit quantizer at the receiver. The inputs to the channel are subject to average power constraints. We show that any point on the boundary of the capacity region is achieved by discrete input distributions. Based on the slope of the tangent line to the capacity region at a boundary point, we propose upper bounds on the cardinality of the support of these distributions.
%Then, we focus on the sum capacity of this channel, we derive a closed-form sum capacity expression, and show that it achieved by time division with power control. As a side result of the optimization trick we employ to obtain the cardinality of the input distributions, we also resolve an open problem for a real point-to-point AWGN channel with a $K$-bin ADC front end (symmetric or asymmetric). We show that a discrete input distribution with at most $K$ mass points is sufficient to achieve the point-to-point capacity.

The paper is organized as follows. Section \ref{s2.1} introduces the system model. In Section \ref{pre}, the capacity region of a general two-transmitter memoryless MAC under input average power constraints is investigated. Through an example, it is shown that when there is input average power constraint, it is necessary to consider the capacity region with the auxiliary random variable $U$ in general. The main result of the paper is presented in Section \ref{pre}, and a detailed proof is given in Section \ref{pprr}. Finally, Section \ref{con} concludes the paper.

\textbf{Notations.} Random variables are denoted by capital letters, while their realizations with lower case letters. $F_{X}(x)$ denotes the cumulative distribution function (CDF) of random variable $X$. The conditional probability mass function (pmf) $p_{Y|X_1,X_2}(y|x_1,x_2)$ will be written as $p(y|x_1,x_2)$. For integers $m\leq n$, we have $[m:n]= \{m, m+1,\ldots,n\}$. For $0\leq t\leq 1$, $H_b(t)\triangleq-t\log_2t-(1-t)\log_2(1-t)$ denotes the binary entropy function. The unit-step function is denoted by $s(\cdot)$.
\section{System model and preliminaries}\label{s2.1}
\begin{figure}[t]
  \centering
  \includegraphics[scale=0.8]{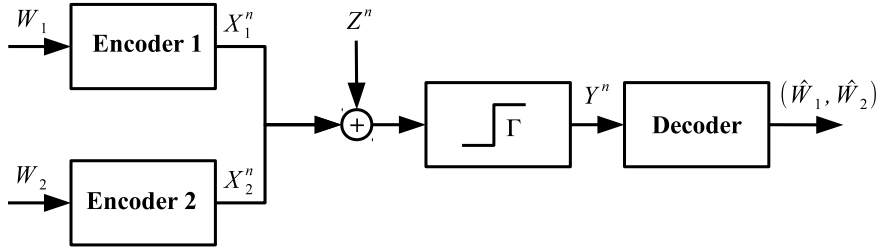}\\
  \caption{A two-transmitter Gaussian MAC with a one-bit ADC front end at the receiver.}\label{fig1}
\end{figure}

We consider a two-transmitter memoryless Gaussian MAC (as shown in Figure \ref{fig1}) with a one-bit quantizer $\Gamma$ at the receiver front end. Transmitter $j = 1,2$ encodes its message $W_j$ into a codeword $X_j^n$ and transmits it over
the shared channel. The signal received by the decoder is given by
\begin{equation*}
Y = \Gamma(X_{1,i}+X_{2,i}+Z_i),\ i\in[1:n],
\end{equation*}
where $\{Z_i\}_{i=1}^n$ is an independent and identically distributed (i.i.d.) Gaussian noise process, also independent of the channel inputs $X_1^n$ and $X_2^n$ with $Z_i\sim \mathcal{N}(0,1), i\in[1:n]$. $\Gamma$ represents the one-bit ADC operation given by
\begin{equation*}
    \Gamma(x)=\left\{\begin{array}{cc} 1 & x\geq 0\\ 0 & x<0 \end{array}\right..
\end{equation*}
This channel can be modelled by the triplet $\left(\mathcal{X}_1\times\mathcal{X}_2,p(y|x_1,x_2),\mathcal{Y}\right)$, where $\mathcal{X}_1,\mathcal{X}_2$ ($=\mathbb{R}$) and $\mathcal{Y}$ ($=\{0,1\}$), respectively, are the alphabets of the inputs and the output. The conditional pmf of the channel output $Y$ conditioned on the channel inputs $X_1$ and $X_2$ (i.e. $p(y|x_1,x_2)$) is characterized by
\begin{equation}\label{ch}
p(0|x_1,x_2) = 1 - p(1|x_1,x_2) = Q(x_1+x_2),
\end{equation}
where $Q(x)\triangleq\frac{1}{\sqrt{2\pi}}\int_{x}^{+\infty}e^{-\frac{t^2}{2}}dt$.

Upon receiving the sequence $Y^n$, the decoder finds the estimates $(\hat{W}_1,\hat{W}_2)$ of the messages.

A $(2^{nR_1},2^{nR_2},n)$ code for this channel consists of (as in \cite{network_info}):
\begin{itemize}
\item two message sets $[1:2^{nR_1}]$ and $[1:2^{nR_2}]$,
\item two encoders, where encoder $j =1,2$ assigns a codeword $x_j^n(w_j)$ to each message $w_j\in [1:2^{nR_j}]$, and
\item a decoder that assigns estimates $(\hat{w}_1,\hat{w}_2)\in [1:2^{nR_1}]\times [1:2^{nR_2}]$ or an error message to each received sequence $y^n$.
\end{itemize}
We assume that the message pair $(W_1,W_2)$ is uniformly distributed over $[1:2^{nR_1}]\times [1:2^{nR_2}]$. The average probability of error is defined as
\begin{equation}
P_e^{(n)}=\mbox{Pr}\left\{(\hat{W}_1,\hat{W}_2)\neq(W_1,W_2)\right\}
\end{equation}
Average power constraints are imposed on the channel inputs as
\begin{equation}\label{pcons}
\frac{1}{n}\sum_{i = 1}^{n}x^2_{j,i}(w_j)\leq P_j\ , \ \forall w_j\in[1:2^{nR_j}],\ j\in\{1,2\},
\end{equation}
where $x_{j,i}(w_j)$ denotes the $i^{\mbox{th}}$ element of the codeword $x_j^n(w_j)$.

A rate pair $(R_1,R_2)$ is said to be \textit{achievable} for this channel if there exists a sequence of $(2^{nR_1},2^{nR_2},n)$ codes (satisfying the average power constraints (\ref{pcons})) such that $\lim_{n\to\infty}P_e^{(n)}=0$. The \textit{capacity region} $\mathscr{C}(P_1,P_2)$ of this channel is
the closure of the set of achievable rate pairs $(R_1,R_2)$.

\section{Main results}\label{pre}
\textbf{Proposition 1.} The capacity region $\mathscr{C}(P_1,P_2)$ of a two-transmitter memoryless MAC with average power constraints $P_1$ and $P_2$ is the set of non-negative rate pairs $(R_1,R_2)$ that satisfy
\begin{align}
R_1&\leq I(X_1;Y|X_2,U),\nonumber\\
R_2&\leq I(X_2;Y|X_1,U),\nonumber\\
R_1+R_2&\leq I(X_1,X_2;Y|U),\label{c1}
\end{align}
for some $F_U(u)F_{X_1|U}(x_1|u)F_{X_2|U}(x_2|u)$, such that $\mathds{E}[X_j^2]\leq P_j,\  j=1,2.$ Also, it is sufficient to consider $|\mathcal{U}|\leq 5$.
%It is obvious that for a fixed joint distribution $F_U(u)F_{X_1|U}(x_1|u)F_{X_2|U}(x_2|u)$, the region in (\ref{c1}) is a pentagon.
\begin{proof}
The capacity region of the discrete memoryless (DM) MAC with input cost constraints has been addressed in Exercise 4.8 of \cite{network_info}. If the input alphabets are not discrete, the capacity region is still the same because: 1) the converse remains the same if the inputs are from a continuous alphabet; 2) the region is achievable by coded time sharing and the discretization procedure (see Remark 3.8 in \cite{network_info}). Therefore, it is sufficient to show the cardinality bound $|\mathcal{U}|\leq 5$.

Let $\mathscr{P}$ be the set of all product distributions (i.e., of the form $F_{X_1}(x_1)F_{X_2}(x_2)$) on $\mathbb{R}^2$. Let $\mathbf g :\mathscr{P}\to\mathbb{R}^5$ be a vector-valued mapping defined element-wise as
\begin{align}
g_1(F_{X_1|U}(\cdot|u)F_{X_2|U}(\cdot|u))&=I(X_1;Y|X_2,U=u),\nonumber\\
g_2(F_{X_1|U}(\cdot|u)F_{X_2|U}(\cdot|u))&=I(X_2;Y|X_1,U=u),\nonumber\\
g_3(F_{X_1|U}(\cdot|u)F_{X_2|U}(\cdot|u))&=I(X_1,X_2;Y|U=u),\nonumber\\
g_4(F_{X_1|U}(\cdot|u)F_{X_2|U}(\cdot|u))&=\mathds{E}[X_1^2|U=u],\nonumber\\
g_5(F_{X_1|U}(\cdot|u)F_{X_2|U}(\cdot|u))&=\mathds{E}[X_2^2|U=u].\label{cara}
\end{align}
Let $\mathscr{G}\subset \mathbb{R}^5$ be the image of $\mathscr{P}$ under the mapping $\mathbf g$ (i.e., $\mathscr{G}=\mathbf g(\mathscr{P})$).
%Note that although $\mathbf g$ is a continuous mapping, $\mathscr{G}$ is not necessarily compact\footnote{Note that compactness is not a necessary condition here, as stated in the footnote on page 267 of \cite{Witsenhausen}.} (since $g_4,g_5$ cannot be bounded).
Given an arbitrary $(U,X_1,X_2)\sim F_UF_{X_1|U}F_{X_2|U}$, we obtain the vector $\mathbf r$ as
\begin{align*}
r_1&=I(X_1;Y|X_2,U)=\int_{\mathcal{U}}I(X_1;Y|X_2,U=u)dF_U(u),\\
r_2&=I(X_2;Y|X_1,U)=\int_{\mathcal{U}}I(X_2;Y|X_1,U=u)dF_U(u),\\
r_3&=I(X_1,X_2;Y|U)=\int_{\mathcal{U}}I(X_1,X_2;Y|U=u)dF_U(u),\\
r_4&=\mathds{E}[X_1^2]=\int_{\mathcal{U}}\mathds{E}[X_1^2|U=u]dF_U(u),\\
r_5&=\mathds{E}[X_2^2]=\int_{\mathcal{U}}\mathds{E}[X_2^2|U=u]dF_U(u).
\end{align*}
Therefore, $\mathbf r$ is in the convex hull of $\mathscr{G}\subset \mathbb{R}^5$. {\color{black}By Carath\'{e}odory's theorem \cite{Witsenhausen}, $\mathbf r$ can be written as a convex combination of 6 ($=5+1$) or fewer points in $\mathscr{G}$}, which states that it is sufficient to consider $|\mathcal{U}|\leq 6$. Since $\mathscr{P}$ is a connected set\footnote{$\mathscr{P}$ is the product of two connected sets, therefore, it is connected. Each of the sets in this product is connected because of being a convex vector space.} and the mapping $\mathbf g$ is continuous\footnote{This is a direct result of the continuity of the channel transition probability.}, $\mathscr{G}$ is a connected subset of $\mathbb{R}^5$. Therefore, connectedness of $\mathscr{G}$ refines the cardinality of $U$ to $|\mathcal{U}|\leq 5$.\footnote{{\color{black}This refinement of the cardinality is due to the connected version of Carath\'{e}odory's theorem as mentioned in \cite[p.267]{Witsenhausen}, which is originally due to \cite[p.35-36]{Eggleston}.}}
\end{proof}
\textbf{Lemma 1.} For the boundary points of $\mathscr{C}(P_1,P_2)$ that are not sum-rate optimal, it is sufficient to have $|\mathcal{U}|\leq 4$.
\begin{proof}
Any point on the boundary of the capacity region that does not maximize $R_1+R_2$, is either of the form $(I(X_1;Y|X_2,U),I(X_2;Y|U))$ or $(I(X_1;Y|U),I(X_2;Y|X_1,U))$ for some $F_UF_{X_1|U}F_{X_2|U}$ that satisfies $\mathds{E}[X_j^2]\leq P_j, j=1,2.$ In other words, it is one of the corner points of the corresponding pentagon in (\ref{c1}). As in the proof of Proposition 1, define the mapping $\mathbf g :\mathscr{P}\to\mathbb{R}^4$, where $g_1$ and $g_2$ are the coordinates of this boundary point conditioned on $U=u$, and $g_3$, $g_4$ are the same as $g_4$ and $g_5$ in (\ref{cara}), respectively. The sufficiency of $|\mathcal{U}|\leq 4$ in this case follows similarly to the proof of Proposition 1.
\end{proof}
When there is no input cost constraint, the capacity region of the MAC can be characterized either through the convex hull operation as in \cite[Theorem 4.2]{network_info}, or with the introduction of an auxiliary random variable as in \cite[Theorem 4.3]{network_info}. {\color{black}The following remark states that when there are input cost constraints, the capacity characterization region requires an auxiliary random variable in general.}

\textbf{Remark 2.} Let $(X_1,X_2)\sim F_{X_1}(x_1)F_{X_2}(x_2)$, such that $\mathds{E}[X_j^2]\leq P_j, j = 1,2$. Let $\mathscr{R}(P_1,P_2)$ denote the set of non-negative rate pairs $(R_1,R_2)$ such that
\begin{align*}
R_1&\leq I(X_1;Y|X_2),\\
R_2&\leq I(X_2;Y|X_1),\\
R_1+R_2&\leq I(X_1,X_2;Y).
\end{align*}
Let $\mathscr{R}_1(P_1,P_2)$ be the convex closure of $\bigcup_{F_{X_1}F_{X_2}}\mathscr{R}(P_1,P_2)$, where the union is over all product distributions that satisfy the average power constraints.

Let $\mathscr{R}_2(P_1,P_2)$ be the set of non-negative rate pairs $(R_1,R_2)$ such that
\begin{align*}
R_1&\leq I(X_1;Y|X_2,U),\\
R_2&\leq I(X_2;Y|X_1,U),\\
R_1+R_2&\leq I(X_1,X_2;Y|U),
\end{align*}
for some $F_U(u)F_{X_1|U}(x_1|u)F_{X_2|U}(x_2|u)$ that satisfies $\mathds{E}[X_j^2|U=u]\leq P_j, j=1,2,\ \forall u$.

It can be verified that $\mathscr{R}_1(P_1,P_2)=\mathscr{R}_2(P_1,P_2)$. By comparing $\mathscr{R}_2(P_1,P_2)$ to the capacity region $\mathscr{C}(P_1,P_2)$, we can conclude that $\mathscr{R}_2(P_1,P_2)\subseteq\mathscr{C}(P_1,P_2)$. This follows from the fact that in the region $\mathscr{R}_2(P_1,P_2)$, the average power constraint $\mathds{E}[X_j^2|U=u]\leq P_j$ holds for every realization of the auxiliary random variable $U$, which is a stronger condition than $\mathds{E}[X_j^2]\leq P_j$ used in the capacity region. The following example shows that $\mathscr{R}_1(P_1,P_2)$ and $\mathscr{R}_2(P_1,P_2)$ can be strictly smaller than $\mathscr{C}(P_1,P_2)$.

Consider the same Gaussian MAC with one-bit quantizer at the receiver (as depicted in Figure \ref{fig1}) with the following changes: i) $\mathcal{X}_1=\mathcal{X}_2=\{-\sqrt{2},0,\sqrt{2}\}$, ii) Besides the average power constraints of $P_1=P_2=1$, we also impose the constraint that the inputs should have a zero mean, i.e. $\mathds{E}[X_j]=0,\ j=1,2.$
%as $\frac{1}{n}\sum_{i = 1}^{n}x_{j,i}(m_j)=0\ , \ \forall m_j\in[1:2^{nR_j}], j\in\{1,2\}$.
The capacity region of this channel is the set of non-negative rate pairs $(R_1,R_2)$ such that (\ref{c1}) holds for some $F_U(u)F_{X_1|U}(x_1|u)F_{X_2|U}(x_2|u)$ which satisfies $\mathds{E}[X_j^2]\leq P_j,\ \mathds{E}[X_j]=0,\  j=1,2.$ Also, let $\mathscr{R}_1$ be the rate region in Remark 2 with the additional constraints $\mathds{E}[X_j]=0,\  j=1,2.$

In order to show that $\mathscr{R}_1$ can be strictly smaller than the capacity region,  we show that there exists a point in the capacity region which is not in $\mathscr{R}_1$. We have,
\begin{align}
\max_{(R_1,R_2)\in\mathscr{R}_1}R_1+R_2&=\max_{\substack{F_{X_1}F_{X_2}: \\ \mathds{E}[X_j^2]\leq 1,\mathds{E}[X_j]=0,\ j=1,2}}I(X_1,X_2;Y)\nonumber\\&=\max_{\substack{F_{X_1}F_{X_2}: \\ \mathds{E}[X_j^2]\leq 1,\mathds{E}[X_j]=0,\ j=1,2}}I(X_1+X_2;Y)\label{c2}\\&\leq\max_{\substack{F_X(x):\\ \mathds{E}[X^2]\leq 2,\mathds{E}[X]=0}}I(X;Y)\label{c3}\\&\leq\max_{\substack{F_X(x):\\ \mathds{E}[X^2]\leq 2}}I(X;Y)\nonumber\\&=1-H_b(Q(\sqrt{2}))\label{c4}\\&\leq \max_{\substack{F_UF_{X_1|U}F_{X_2|U}: \\ \mathds{E}[X_j^2]\leq 1,\mathds{E}[X_j]=0,\ j=1,2}}I(X_1,X_2;Y|U)\label{c6}\\&=\max_{(R_1,R_2)\in\mathscr{C}}R_1+R_2\nonumber,
\end{align}
where (\ref{c2}) is due to the fact that $X_1+X_2$ is a function of the pair $(X_1,X_2)$, and the following Markov chain holds: $(X_1,X_2)\to X_1+X_2\to Y$. In (\ref{c3}), we use the inequality $\mathds{E}[(X_1+X_2)^2]=\mathds{E}[X_1^2]+\mathds{E}[X_2^2]\leq 2$, since $X_1$ and $X_2$ are independent and zero mean. Also, the channel from $X$ to $Y$ is characterized by the conditional distribution $p_{Y|X}(y|x)\sim\mbox{Bern}(Q(x))$. Equality in (\ref{c4}) is due to \cite{jaspreet}, where the maximum is shown to be achieved by the CDF $F^*_X(x)=\frac{1}{2}s(x+\sqrt{2})+\frac{1}{2}s(x-\sqrt{2})$, where $s(\cdot)$ is the unit step function. Let $U\sim\mbox{Bern}(\frac{1}{2})$, $F_{X_1|U}(x|1)=F_{X_2|U}(x|0)=\frac{1}{2}s(x+\sqrt{2})+\frac{1}{2}s(x-\sqrt{2})$ and $F_{X_1|U}(x|0)=F_{X_2|U}(x|1)=s(x)$. For this joint distribution on $(U,X_1,X_2)$, we have $\mathds{E}[X_j]=0,\ \mathds{E}[X_j^2]\leq 1,\ j=1,2$, and $I(X_1,X_2;Y|U)=1-H_b(Q(\sqrt{2}))$, which results in (\ref{c6}).

\begin{figure}[t]
  \centering
  \includegraphics[width=13cm]{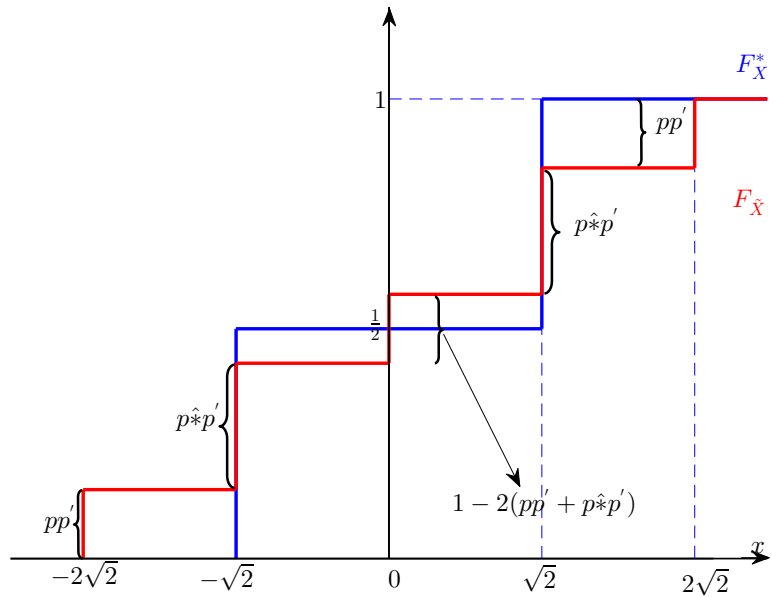}\\
  \caption{The distributions of $X^*$ and $\tilde{X}$.}\label{fig2}
\end{figure}

In what follows, it is proved that the inequality in (\ref{c3}) is strict. In other words, the sum rate of $1-H_b(Q(\sqrt{2}))$ cannot be obtained by any rate pair in $\mathscr{R}_1$, while it belongs to the capacity region. Let $\tilde{X}=X_1+X_2$, where $X_1$ and $X_2$ are two zero-mean independent random variables on $\mathcal{X}_1$($=\mathcal{X}_2$) satisfying the average power constraint $\mathds{E}[X_j^2]\leq 1, j=1,2$. We show that the minimum L\'{e}vy distance\footnote{The L\'{e}vy distance between two distributions $F,G:\mathbb{R}\to[0,1]$ is defined as
\begin{equation*}
d_L(F,G)=\inf\{\epsilon>0|F(x-\epsilon)-\epsilon\leq G(x)\leq F(x+\epsilon)+\epsilon,\ \forall x\in\mathbb{R}\}.
\end{equation*}} between $F^*_X(x)$ and all the distributions $F_{\tilde{X}}(x)$ (induced by $F_{X_1}F_{X_2}$) is bounded away from zero. Since $\mathds{E}[X_j]=0$ and $\mathds{E}[X_j^2]\leq 1,\ j=1,2$, the distribution of $X_1$ is
$F_{X_1}(x)=ps(x+\sqrt{2})+(1-2p)s(x) + ps(x-\sqrt{2})$ with $p\leq \frac{1}{4}$. The same applies to $F_{X_2}$ with parameter $p'$($\leq \frac{1}{4}$). The distribution of $\tilde{X}$ induced by $F_{X_1}F_{X_2}$ is given by
\begin{equation*}
F_{\tilde{X}}(x)=pp's(x+2\sqrt{2})+p\hat{*}p's(x+\sqrt{2})+\left(1-2(pp'+p\hat{*}p')\right)s(x)+p\hat{*}p's(x-\sqrt{2})+pp's(x-2\sqrt{2}),
\end{equation*}
where $p\hat{*}p'\triangleq p(1-2p')+p'(1-2p)$ is similar to convolution operation. Let $\tilde{\mathscr{F}}$ be the set of all distributions on $\tilde{X}$ obtained in this way. It can be easily verified that (see Figure \ref{fig2}) for any given $p,p'\leq \frac{1}{4}$, the L\'{e}vy distance between $F_{\tilde{X}}$ and $F^*_X$ is
\begin{equation}
d_L(F_{\tilde{X}},F^*_X)=\max\left\{pp',\frac{1}{2}-pp'-p\hat{*}p'\right\}=\frac{1}{2}-pp'-p\hat{*}p'.
\end{equation}
Subsequently,
\begin{equation*}
\min_{p,p'\leq\frac{1}{4}}d_L(F_{\tilde{X}},F^*_X)=\frac{3}{16}.
\end{equation*}
This shows that there is a neighborhood of $F^*_X$ whose intersection with $\tilde{\mathscr{F}}$ is empty. Note that any neighborhood with radius less than $\frac{3}{16}$ has this property. Combined with the facts that the mutual information is continuous and $F^*_X$ is the unique solution\footnote{This is due to the strict convexity of $H_b(Q(\sqrt{\cdot}))$, which is used in Jensen's inequality in \cite{jaspreet}.}, it proves that the inequality in (\ref{c3}) is strict. Therefore, $\mathscr{R}_1$ ($=\mathscr{R}_2$) is smaller than the capacity region in general.

The main result of this paper is provided in the following theorem. It bounds the cardinality of the support set of the capacity achieving distributions.

\textbf{Theorem 1.} Let $P$ be an arbitrary point on the boundary of the capacity region $\mathscr{C}(P_1,P_2)$ of the memoryless MAC with a one-bit ADC front end\footnote{The results remain valid if the one-bit ADC has a non-zero threshold.} (as shown in Figure \ref{fig1}). {\color{black}$P$ is achieved by a distribution in the form of $F^P_U(u)F^P_{X_1|U}(x_1|u)F^P_{X_2|U}(x_2|u)$.} Also, let $l_P$ be the slope of the line tangent to the capacity region at this point. For any $u\in\mathcal{U}$, the conditional input distributions $F^P_{X_1|U}(x_1|u)$ and {\color{black}$F^P_{X_2|U}(x_2|u)$} have at most $n_1$ and $n_2$ points of increase\footnote{A point $Z$ is said to be a point of increase of a distribution if for any open set $\Omega$ containing $Z$, we have $\mbox{Pr}\{\Omega\}>0.$}, respectively, where
{\color{black}
\begin{equation}\label{Mor1}
(n_1,n_2)=\left\{\begin{array}{cc} (3,5) & l_p<-1\\ (3,3) & l_p=-1\\ (5,3) & l_p>-1 \end{array}\right..
\end{equation}}
\begin{proof}
The proof is provided in Section \ref{pprr}.
\end{proof}
Proposition 1, Lemma 1 and Theorem 1 establish upper bounds on the number of mass points of the distributions that achieve a boundary point. The significance of this result is that once it is known that the optimal inputs are discrete with at most certain number of mass points, the capacity region along with the optimal distributions can be obtained via computer programs.
\section{Proof of theorem 1}\label{pprr}
%A limiting argument is necessary to derive $|U|\leq 5.$ More specifically, we can exclude those $F_{U,X_1,X_2}(u,x_1,x_2)$ in which $X_j|{U=u} (j=1,2)$ has unbounded power i.e., $E(X_j^2|u)>M,\ \forall M\in \mathbb{R}.$ This is due to the fact that there is an average power constraint on $X_1,X_2$. Now, assume that
{\color{black}In order to show that the boundary points of the capacity region are achieved, it is sufficient to show that the capacity region is a closed set, i.e., it includes all of its limit points.

Let $\mathcal{U}$ be a set with $|\mathcal{U}|\leq 5$, and $\Omega$ be defined as
\begin{equation}
\Omega = \big\{F_{U,X_1,X_2}\big|\ U\in\mathcal{U},\ X_1-U-X_2,\ \mathds{E}[X_j^2]\leq P_j,\ j=1,2\big\},
\end{equation}
which is the set of all CDFs on the triplet $(U,X_1,X_2)$, where $U$ is drawn from $\mathcal{U}$, and the Markove chain $X_1-U-X_2$ and the corresponding average power constraints hold.

In Appendix \ref{appp1}, it is proved that $\Omega$ is a compact set. Since a continuous mapping preserves compactness, the capacity region is compact. Since the capacity region is a subset of $\mathds{R}^2$, it is closed and bounded\footnote{Note that a subset of $\mathds{R}^k$ is compact if and only if it is closed and bounded \cite{rudin}.}. Therefore, any point $P$ on the boundary of the capacity region is achieved by a distribution denoted by $F^P_U(u)F^P_{X_1|U}(x_1|u)F^P_{X_2|U}(x_2|u)$.}

Since the capacity region is a convex space, it can be characterized by its supporting hyperplanes. In other words, any point on the boundary of the capacity region, denoted by $(R_1^b,R_2^b)$, can be written as
\begin{equation*}
(R_1^b,R_2^b)=\mbox{arg}\max_{(R_1,R_2)\in\mathscr{C}(P_1,P_2)}R_1+\lambda R_2,
\end{equation*}
for some $\lambda>0$.

Any rate pair $(R_1,R_2)\in\mathscr{C}(P_1,P_2)$ must lie within a pentagon defined by (\ref{c1}) for some $F_UF_{X_1|U}F_{X_2|U}$ that satisfies the power constraints. Therefore, due to the structure of the pentagon, the problem of finding the boundary points is equivalent to the following maximization problem.
\begin{equation}\label{harchi}
\max_{(R_1,R_2)\in\mathscr{C}(P_1,P_2)}R_1+\lambda R_2=\left\{\begin{array}{cc} \max I(X_1;Y|X_2,U) + \lambda I(X_2;Y|U) & 0<\lambda\leq1\\  \max I(X_2;Y|X_1,U) + \lambda I(X_1;Y|U) & \lambda>1 \end{array}\right.,
\end{equation}
where on the right hand side (RHS) of (\ref{harchi}), the maximizations are over all $F_UF_{X_1|U}F_{X_2|U}$ that satisfy the power constraints. It is obvious that when $\lambda = 1$, the two lines in (\ref{harchi}) are the same, which results in the sum capacity.

For any product of distributions $F_{X_1}F_{X_2}$ and the channel in (\ref{ch}), let $I_\lambda$ be defined as
\begin{equation}\label{lambdadef}
I_\lambda(F_{X_1}F_{X_2}) = \left\{\begin{array}{cc} I(X_1;Y|X_2) + \lambda I(X_2;Y) & 0<\lambda\leq 1\\ I(X_2;Y|X_1) + \lambda I(X_1;Y) & \lambda>1 \end{array}\right..
\end{equation}
With this definition, (\ref{harchi}) can be rewritten as
\begin{equation*}
\max_{(R_1,R_2)\in\mathscr{C}(P_1,P_2)}R_1+\lambda R_2=\max\sum_{i=1}^{5}p_U(u_i)I_\lambda(F_{X_1|U}(x_1|u_i)F_{X_2|U}(x_2|u_i)),
\end{equation*}
where the second maximization is over product distributions of the form $p_U(u)F_{X_1|U}(x_1|u)F_{X_2|U}(x_2|u),\ |\mathcal{U}|\leq 5$, such that
\begin{equation*}
\sum_{i=1}^{5}p_U(u_i)\mathds{E}[X_j^2|U=u_i]\leq P_j,\ j=1,2.
\end{equation*}

\textbf{Proposition 2.} For a given $F_{X_1}$ and any $\lambda>0$, $I_\lambda(F_{X_1}F_{X_2})$ is a concave, continuous and weakly differentiable function of $F_{X_2}$. In the statement of this Proposition, $F_{X_1}$ and $F_{X_2}$ could be interchanged.
\begin{proof}
The proof is provided in Appendix \ref{app1}.
\end{proof}

\textbf{Proposition 3.} Let $P_1',P_2'$ be two arbitrary non-negative real numbers. For the following problem
\begin{equation}
\max_{\substack{F_{X_1}F_{X_2}: \\ \mathds{E}[X_j^2]\leq P_j',\ j=1,2}} I_\lambda(F_{X_1}F_{X_2}),
\end{equation}
the optimal inputs $F_{X_1}^*$ and $F_{X_2}^*$, which are not unique in general, have the following properties,
\begin{enumerate}[label=(\roman*)]
\item The support sets  of $F_{X_1}^*$ and $F_{X_2}^*$ are bounded subsets of $\mathbb{R}$.
\item $F_{X_1}^*$ and $F_{X_2}^*$ are discrete distributions that have at most $n_1$ and $n_2$ points of increase, respectively, where
{\color{black}
\begin{equation*}
(n_1,n_2)=\left\{\begin{array}{cc} (5,3) & 0<\lambda<1\\ (3,3) & \lambda=1\\ (3,5) & \lambda>1 \end{array}\right..
\end{equation*}}
\end{enumerate}
\begin{proof}
%First, assume $0<\lambda<1$. We show that given $X_1\sim F_{X_1}$, the distribution $F_{X_2}^*$ that maximizes $I_\lambda$ has at most 3 points of increase. Afterwards, we show that given any $X_2\sim F_{X_2}$ with at most 3 mass points, the distribution $F_{X_1}^*$ that maximizes $I_\lambda$ has at most 5 points of increase.
We start with the proof of the first claim. Assume that $0<\lambda\leq 1$, and $F_{X_2}$ is given. Consider the following optimization problem:
\begin{equation}
I^*_{F_{X_2}}\triangleq \sup_{\substack{F_{X_1}: \\ \mathds{E}[X_1^2]\leq P_1'}}I_{\lambda}(F_{X_1}F_{X_2}).\label{op1}
\end{equation}
Note that $I^*_{F_{X_2}}<+\infty$, since for any $\lambda>0$, from (\ref{lambdadef}),
\begin{equation*}
I_\lambda\leq (\lambda+1)H(Y)\leq (1+\lambda)<+\infty.
\end{equation*}
From Proposition 2, $I_\lambda$ is a continuous, concave function of $F_{X_1}$. Also, the set of all CDFs with bounded second moment (here, $P_1'$) is convex and compact\footnote{The compactness follows from \cite[Appendix I]{Abou}. The only difference is in using Chebyshev's inequality instead of Markov inequality.}. Therefore, the supremum in (\ref{op1}) is achieved by a distribution $F_{X_1}^*$. Since for any $F_{X_1}(x)=s(x-x_0)$ with $|x_0|^2<P_1'$, we have $\mathds{E}[X_1^2]<P_1'$, the Lagrangian theorem and the Karush-Kuhn-Tucker conditions state that there exists a $\theta_1\geq 0$ such that
\begin{align}
I^*_{F_{X_2}}&=\sup_{F_{X_1}}\left\{I_\lambda(F_{X_1}F_{X_2})-\theta_1\left(\int x^2dF_{X_1}(x)-P_1'\right)\right\}\label{yekom}.
\end{align}
Furthermore, the supremum in (\ref{yekom}) is achieved by $F_{X_1}^*$, and
\begin{equation}\label{sefr}
\theta_1\left(\int x^2dF_{X_1}^*(x)-P_1'\right)=0.
\end{equation}
%
%
%\begin{align}
%I^*_{F_{X_1}}&=\max_{F_{X_2}}\left\{I_\lambda(F_{X_1}F_{X_2})-\theta_2\left(\int x^2dF_{X_2}(x)-P_2'\right)\right\}\label{yekom'}\\&=I_\lambda(F_{X_1}F_{X_2}^*)-\theta_2\left(\int x^2dF_{X_2}^*(x)-P_2'\right),
%\end{align}
%and $\theta_2\left(\int x^2dF_{X_2}^*(x)-P_2'\right)=0$.

\textbf{Lemma 2.} The Lagrangian multiplier $\theta_1$ is nonzero.
\begin{proof}
Having a zero Lagrangian multiplier means the power constraint is inactive. In other words, if $\theta_1=0$, (\ref{op1}) and (\ref{yekom}) imply that
\begin{equation}\label{khalaf}
\sup_{\substack{F_{X_1} \\ \mathds{E}[X_1^2]\leq P_1'}}I_{\lambda}(F_{X_1}F_{X_2})=\sup_{F_{X_1} }I_{\lambda}(F_{X_1}F_{X_2}).
\end{equation}
We prove that (\ref{khalaf}) does not hold by showing that its left hand side (LHS) is strictly less than 1, while its RHS equals 1. The details are provided in Appendix \ref{app2}.
\end{proof}
$I_{\lambda}(F_{X_1}F_{X_2})$ ($0<\lambda\leq 1$) can be written as
\begin{align}
I_\lambda(F_{X_1}F_{X_2}) &=\int_{-\infty}^{+\infty}\int_{-\infty}^{+\infty}\sum_{y=0}^{1}p(y|x_1,x_2)\log\frac{p(y|x_1,x_2)}{[p(y;F_{X_1}F_{X_2})]^\lambda[p(y;F_{X_1}|x_2)]^{1-\lambda}}dF_{X_1}(x_1)dF_{X_2}(x_2)\nonumber\\
& = \int_{-\infty}^{+\infty}\tilde{i}_\lambda(x_1;F_{X_1}|F_{X_2})dF_{X_1}(x_1)\label{mar1}\\
& =\int_{-\infty}^{+\infty}i_\lambda(x_2;F_{X_2}|F_{X_1})dF_{X_2}(x_2),
\end{align}
%The notation $p(y;F_{X_1}F_{X_2})$ emphasizes the fact that the pmf of $Y$ has been induced by the product $F_{X_1}F_{X_2}$. Similarly, $p(y;F_{X_1}|x_2)$ shows that the conditional pmf of $Y$ conditioned on $x_2$ has been induced by $F_{X_1}$ (i.e., $p(y;F_{X_1}|x_2)=\int p(y|x_1,x_2)dF_{X_1}$).
where we have defined
\begin{equation}
\tilde{i}_\lambda(x_1;F_{X_1}|F_{X_2})\triangleq\int_{-\infty}^{+\infty}\left(D\left(p(y|x_1,x_2)||p(y;F_{X_1}F_{X_2})\right)+(1-\lambda)\sum_{y=0}^{1}p(y|x_1,x_2)\log\frac{p(y;F_{X_1}F_{X_2})}{p(y;F_{X_1}|x_2)}\right)dF_{X_2}(x_2),\label{dden1}
\end{equation}
and
\begin{equation}
i_\lambda(x_2;F_{X_2}|F_{X_1})\triangleq\int_{-\infty}^{+\infty}D\left(p(y|x_1,x_2)||p(y;F_{X_1}F_{X_2})\right)dF_{X_1}(x_1)-(1-\lambda)D\left(p(y;F_{X_1}|x_2)||p(y;F_{X_1}F_{X_2})\right).\label{dden2}
\end{equation}
$p(y;F_{X_1}F_{X_2})$ is nothing but the pmf of $Y$ with the emphasis that it has been induced by $F_{X_1}$ and $F_{X_2}$. Likewise, $p(y;F_{X_1}|x_2)$ is the conditional pmf $p(y|x_2)$ when $X_1$ is drawn according to $F_{X_1}$.
From (\ref{mar1}), $\tilde{i}_\lambda(x_1;F_{X_1}|F_{X_2})$ can be considered as the density of $I_\lambda$ over $F_{X_1}$ when $F_{X_2}$ is given. $i_\lambda(x_2;F_{X_2}|F_{X_1})$ can be interpreted in a similar way.

Note that (\ref{yekom}) is an unconstrained optimization problem over the set of all CDFs. Since $\int x^2dF_{X_1}(x)$ is linear and weakly differentiable in $F_{X_1}$, the objective function in (\ref{yekom}) is {\color{black}concave} and weakly differentiable. Hence, a necessary condition for optimality of $F_{X_1}^*$ is
\begin{equation}\label{nec}
\int\{\tilde{i}_\lambda(x_1;F_{X_1}^*|F_{X_2})+\theta_1(P_1'-x_1^2)\}dF_{X_1}(x_1)\leq I^*_{F_{X_2}},\ \ \forall F_{X_1}.
\end{equation}
%where $\tilde{i}_\lambda(x_1;F_{X_1}|F_{X_2})$ is the density of $I_\lambda$ over $F_{X_1}$ when $F_{X_2}$ is given (see (\ref{dden1})).
{\color{black}Furthermore, (\ref{nec}) can be verified to be equivalent to
\begin{align}
\tilde{i}_\lambda(x_1;F_{X_1}^*|F_{X_2})+\theta_1(P_1'-x_1^2)&\leq I^*_{F_{X_2}},\ \ \forall x_1\in\mathbb{R},\label{nec2}\\
\tilde{i}_\lambda(x_1;F_{X_1}^*|F_{X_2})+\theta_1(P_1'-x_1^2)&= I^*_{F_{X_2}},\ \ \mbox{if }x_1\mbox{ is a point of increase of }F_{X_1}^*.\label{nec3}
\end{align}
The justifications of (\ref{nec}), (\ref{nec2}) and (\ref{nec3}) are provided in Appendix \ref{just}.}

%First we show that $\theta\neq 0.$ Note that when $0<\lambda\leq 1$
%\begin{align}
%\sup_{\substack{F_{X_2} \\ \mathds{E}[X_2^2]\leq P_2'}}I_{\lambda}(F_{X_1}F_{X_2})&=\sup_{\substack{F_{X_2} \\ \mathds{E}[X_2^2]\leq P_2'}}I(X_1;Y|X_2) + \lambda I(X_2;Y)\nonumber\\
%&\leq \sup_{\substack{F_{X_2} \\ \mathds{E}[X_2^2]\leq P_2'}}I(X_1,X_2;Y)\label{avvali}\\
%&\leq \sup_{\substack{F_{X_1}F_{X_2} \\\mathds{E}[X_1^2]\leq P_1'\\ \mathds{E}[X_2^2]\leq P_2'}}I(X_1+X_2;Y)\nonumber\\
%&\leq \sup_{\mathds{E}[X^2]\leq P_1'+P_2'+2P_1'P_2'}I(X;Y)\nonumber\\
%&= 1-H(Q(\sqrt{P_1'+P_2'+2P_1'P_2'}))\nonumber\\
%&< 1
%\end{align}
In what follows, we prove that in order to satisfy (\ref{nec3}), $F_{X_1}^*$ must have a bounded support by showing that the LHS of (\ref{nec3}) goes to $-\infty$ with $x_1$. The following lemma is useful in the sequel for taking the limit processes inside the integrals.

\textbf{Lemma 3.} Let $X_1$ and $X_2$ be two independent random variables satisfying $\mathds{E}[X_1^2]\leq P_1'$ and $\mathds{E}[X_2^2]\leq P_2'$, respectively ($P_1',P_2'\in[0,+\infty)$). Considering the conditional pmf in (\ref{ch}), the following inequalities hold.
\begin{align}
\bigg|D\left(p(y|x_1,x_2)||p(y;F_{X_1}F_{X_2})\right)\bigg|&\leq 1- 2\log Q(\sqrt{P_1'}+\sqrt{P_2'})\label{Le1}\\
p(y;F_{X_1}|x_2)&\geq Q\left(\sqrt{P_1'}+|x_2|\right)\label{Le2}\\
\color{black}\left|\sum_{y=0}^{1}p(y|x_1,x_2)\log\frac{p(y;F_{X_1}F_{X_2})}{p(y;F_{X_1}|x_2)}\right|&\color{black}\leq-2\log Q\left(\sqrt{P_1'}+\sqrt{P_2'}\right)-2\log Q\left(\sqrt{P_1'}+|x_2|\right)\label{Le3}
\end{align}
\begin{proof}
The proof is provided in Appendix \ref{app3}.
\end{proof}
Note that
%\begin{align}
%\lim_{x_1\to+\infty}p_{Y|X_1}(0;F_{X_2}|x_1)&=\lim_{x_1\to+\infty}\int_{-\infty}^{+\infty}Q(x_1+x_2)dF_{X_2}(x_2)\nonumber\\
%&=\int_{-\infty}^{+\infty}\lim_{x_1\to+\infty}Q(x_1+x_2)dF_{X_2}(x_2)\label{leb1}\\
%&=0\label{zz},
%\end{align}
%where (\ref{leb1}) follows from (\ref{yek}). From (\ref{zz}), we can write
%\begin{align}
%\lim_{x_2\to+\infty}D\left(p(y;F_{X_1}|x_2)||p(y;F_{X_1}F_{X_2}^*)\right)&=\lim_{x_2\to+\infty}\sum_{y=0}^{1}p(y;F_{X_1}|x_2)\log\frac{p(y;F_{X_1}|x_2)}{p(y;F_{X_1}F_{X_2}^*)}\nonumber\\
%&=-\log p_Y(1;F_{X_1}F_{X_2}^*)\label{yy}.
%\end{align}
%Also,
\begin{align}
\lim_{x_1\to+\infty}\int_{-\infty}^{+\infty}D\left(p(y|x_1,x_2)||p(y;F_{X_1}^*F_{X_2})\right)dF_{X_2}(x_2)&=\int_{-\infty}^{+\infty}\lim_{x_1\to+\infty}D\left(p(y|x_1,x_2)||p(y;F_{X_1}^*F_{X_2})\right)dF_{X_2}(x_2)\label{ww}\\
&= -\log p_Y(1;F_{X_1}^*F_{X_2})\label{www'}\\
&\leq -\log Q(\sqrt{P_1'}+\sqrt{P_2'})\label{ww1},
\end{align}
where (\ref{ww}) is due to Lebesgue dominated convergence theorem \cite{rudin} and (\ref{Le1}), which permit the interchange of the limit and the integral; {\color{black}(\ref{www'}) is due to the following
\begin{align*}
\lim_{x_1\to+\infty}D\left(p(y|x_1,x_2)||p(y;F_{X_1}^*F_{X_2})\right)&=\lim_{x_1\to+\infty}\sum_{y=0}^{1}p(y|x_1,x_2)\log\frac{p(y|x_1,x_2)}{p(y;F_{X_1}^*F_{X_2})}\nonumber\\
&=-\log p_Y(1;F_{X_1}^*F_{X_2}),
\end{align*}
since $p(0|x_1,x_2)=Q(x_1+x_2)$ goes to zero when $x_1\to +\infty$ and $p_Y(y;F_{X_1}^*F_{X_2})\ (y=0,1)$ is bounded away from zero by (\ref{vv})}
; and (\ref{ww1}) is obtained from (\ref{vv}) in Appendix \ref{app3}. Furthermore,
\begin{align}
\lim_{x_1\to+\infty}\int_{-\infty}^{+\infty}\sum_{y=0}^{1}p(y|x_1,x_2)\log\frac{p(y;F_{X_1}^*F_{X_2})}{p(y;F_{X_1}^*|x_2)}dF_{X_2}(x_2)&=\int_{-\infty}^{+\infty}\!\!\lim_{x_1\to+\infty}\sum_{y=0}^{1}p(y|x_1,x_2)\log\frac{p(y;F_{X_1}^*F_{X_2})}{p(y;F_{X_1}^*|x_2)}dF_{X_2}(x_2)\label{ped1}\\&=\log p_Y(1;F_{X_1}^*F_{X_2})-\int_{-\infty}^{+\infty}\log p(1;F_{X_1}^*|x_2)dF_{X_2}(x_2)\nonumber\\
&<-\log Q\left(\sqrt{P_1'}+\sqrt{P_2'}\right)\label{ped2},
\end{align}
where (\ref{ped1}) is due to Lebesgue dominated convergence theorem along with (\ref{Le3}) and (\ref{xcxc}) in Appendix \ref{app3}; (\ref{ped2}) is from (\ref{Le2}) and convexity of $\log Q(\alpha +\sqrt{t})$ in $t$ when $\alpha\geq 0$ (see Appendix \ref{app4}).

Therefore, from (\ref{ww1}) and (\ref{ped2}),
\begin{align}\label{yeki}
\lim_{x_1\to+\infty}\tilde{i}_\lambda(x_1;F_{X_1}^*|F_{X_2})\leq\color{black}-(2-\lambda) \log Q(\sqrt{P_1'}+\sqrt{P_2'})<+\infty.
\end{align}
Using a similar approach, we can also obtain
\begin{align}\label{dota}
\lim_{x_1\to-\infty}\tilde{i}_\lambda(x_1;F_{X_1}^*|F_{X_2})\leq\color{black}-(2-\lambda) \log Q(\sqrt{P_1'}+\sqrt{P_2'})<+\infty.
\end{align}
From (\ref{yeki}), (\ref{dota}) and the fact that $\theta_1>0$ (see Lemma 2), the LHS of (\ref{nec2}) goes to $-\infty$ when $|x_1|\to +\infty$. Since any point of increase of $F_{X_1}^*$ must satisfy (\ref{nec2}) with equality, and $I_{F_{X_2}}^*\geq0$, it is proved that $F_{X_1}^*$ has a bounded support, i.e., $X_1\in [A_1,A_2]$ for some $A_1,A_2\in\mathbb{R}$.\footnote{Note that $A_1$ and $A_2$ are determined by the choice of $F_{X_2}$.}

Similarly, for a given $F_{X_1}$, the optimization problem
\begin{equation*}
I^*_{F_{X_1}}=\sup_{\substack{F_{X_2}: \\ \mathds{E}[X_2^2]\leq P_2'}}I_{\lambda}(F_{X_1}F_{X_2})\label{orp1},
\end{equation*}
boils down to the following necessary condition
\begin{align}
i_\lambda(x_2;F_{X_2}^*|F_{X_1})+\theta_2(P_2'-x_2^2)&\leq I^*_{F_{X_1}},\ \ \forall x_2\in\mathbb{R},\label{nrec2}\\
i_\lambda(x_2;F_{X_2}^*|F_{X_1})+\theta_2(P_2'-x_2^2)&= I^*_{F_{X_1}},\ \
\mbox{if }x_2 \mbox{ is a point of increase of }F_{X_2}^*,\label{nrec22}
\end{align}
for the optimality of $F_{X_2}^*$. However, there are two main differences between (\ref{nrec22}) and (\ref{nec3}). First is the difference between $i_\lambda$ and $\tilde{i}_\lambda$. Second is the fact that we do not claim $\theta_2$ to be nonzero, since the approach used in Lemma 2 cannot be readily applied to $\theta_2$. Nonetheless, the boundedness of the support of $F_{X_2}^*$ can be proved by inspecting the behaviour of the LHS of (\ref{nrec22}) when $|x_2|\to+\infty$.

{\color{black}In what follows, i.e., from (\ref{sipij}) to (\ref{chels}), we prove that the support of $F^*_{X_2}$ is bounded by showing that (\ref{nrec22}) does not hold when $|x_2|$ is above a certain threshold. The first term on the LHS of (\ref{nrec22}) is $i_\lambda(x_2;F_{X_2}^*|F_{X_1})$. From (\ref{dden2}) and (\ref{Le1}), it can be easily verified that
\begin{align}\label{sipij}
\lim_{x_2\to +\infty}i_\lambda(x_2;F_{X_2}^*|F_{X_1})&=-\lambda\log p_Y(1;F_{X_1}F_{X_2}^*)\leq-\lambda \log Q(\sqrt{P_1'}+\sqrt{P_2'}),\nonumber\\
\lim_{x_2\to -\infty}i_\lambda(x_2;F_{X_2}^*|F_{X_1})&=-\lambda\log p_Y(0;F_{X_1}F_{X_2}^*)\leq-\lambda \log Q(\sqrt{P_1'}+\sqrt{P_2'})
\end{align}
From (\ref{sipij}), if $\theta_2>0$, the LHS of (\ref{nrec22}) goes to $-\infty$ with $|x_2|$, which proves that $X_2^*$ is bounded.

For the possible case of $\theta_2=0$, in order to show that (\ref{nrec22}) does not hold when $|x_2|$ is above a certain threshold, we rely on the boundedness of $X_1$. Note that, since $F_{X_1}$ has a bounded support, we denote its support, without loss of generality,  by $[-A_1,A_2]$, where $A_1,A_2$ are some non-negative real numbers. Then, we prove that $i_\lambda$ approaches its limit in (\ref{sipij}) from below. In other words, there is a real number $K$ such that $i_\lambda(x_2;F_{X_2}^*|F_{X_1})<-\lambda\log p_Y(1;F_{X_1}F_{X_2}^*)$ when $x_2>K$, and $i_\lambda(x_2;F_{X_2}^*|F_{X_1})<-\lambda\log p_Y(0;F_{X_1}F_{X_2}^*)$ when $x_2<-K$. This establishes the boundedness of $X_2^*$. In what follows, we only show the former, i.e., when $x_2\to +\infty$. The latter, i.e., $x_2\to -\infty$, follows similarly, and it is omitted for the sake of brevity. }

By rewriting $i_\lambda$, we have
\begin{align}
i_\lambda(x_2;F_{X_2}^*|F_{X_1})&=-\lambda p(1;F_{X_1}|x_2)\log p_Y(1;F_{X_1}F_{X_2}^*)\nonumber\\&\ \ \ -\int_{-A_1}^{A_2}H_b(Q(x_1+x_2))dF_{X_1}(x_1)+(1-\lambda)\!\!\!\!\!\!\!\underbrace{H(Y|X_2=x_2)}_{H_b\left(\int Q(x_1+x_2)dF_{X_1}(x_1)\right)}\!\!\!\!\!\!-\lambda \!\!\!\!\!\overbrace{p(0;F_{X_1}|x_2)}^{\int Q(x_1+x_2)dF_{X_1}(x_1)}\!\!\!\log p_Y(0;F_{X_1}F_{X_2}^*).\label{chehel}
\end{align}
It is obvious that the first term in the RHS of (\ref{chehel}) approaches $-\lambda \log p_Y(1;F_{X_1}F_{X_2}^*)$ from below when $x_2\to+\infty$, since $p(1;F_{X_1}|x_2)\leq 1$. It is also obvious that the remaining terms go to zero when $x_2\to+\infty$. Hence, it is sufficient to show that the second line of (\ref{chehel}) approaches zero from below, which is proved by using the following lemma.

\textbf{Lemma 4.} Let $X_1$ be distributed on $[-A_1,A_2]$ according to $F_{X_1}(x_1)$. We have
\begin{align}
\lim_{x_2\to+\infty}\frac{\int_{-A_1}^{A_2}H_b(Q(x_1+x_2))dF_{X_1}(x_1)}{H_b\bigg(\int_{-A_1}^{A_2} Q(x_1+x_2)dF_{X_1}(x_1)\bigg)}=1.\label{trr}
\end{align}
\begin{proof}
The proof is provided in Appendix \ref{app5}.
\end{proof}
From (\ref{trr}), we can write
\begin{equation}\label{shayad}
\int_{-A_1}^{A_2}H_b(Q(x_1+x_2))dF_{X_1}(x_1)=\gamma(x_2)H_b\left(\int_{-A_1}^{A_2}Q(x_1+x_2)dF_{X_1}(x_1)\right),
\end{equation}
where $\gamma(x_2)\leq 1$ (due to concavity of $H_b(\cdot)$), and $\gamma(x_2)\to 1$ when $x_2\to+\infty$ (due to (\ref{trr})). Also, from the fact that $\lim_{x\to 0}\frac{H_b(x)}{cx}=+\infty\ (c>0)$, we have
\begin{equation}\label{shayad2}
H_b\left(\int_{-A_1}^{A_2}Q(x_1+x_2)dF_{X_1}(x_1)\right)=-\eta(x_2)\log p_Y(0;F_{X_1}F_{X_2}^*)\int_{-A_1}^{A_2}Q(x_1+x_2)dF_{X_1}(x_1),
\end{equation}
where $\eta(x_2)>0$ and $\eta(x_2)\to +\infty$ when $x_2\to+\infty$.
From (\ref{shayad}) and (\ref{shayad2}), the second line of (\ref{chehel}) becomes
\begin{align}\label{chels}
\left(1-\gamma(x_2)+\frac{\lambda}{\eta(x_2)}-\lambda\right)\underbrace{\left(-\eta(x_2)\log p_Y(0;F_{X_1}F_{X_2}^*)\int_{-A_1}^{A_2}Q(x_1+x_2)dF_{X_1}(x_1)\right)}_{\geq 0}.
\end{align}
Since $\gamma(x_2)\to 1$ and $\eta(x_2)\to +\infty$ as $x_2\to+\infty$, there exists a real number $K$ such that $1-\gamma(x_2)+\frac{\lambda}{\eta(x_2)}-\lambda<0$ when $x_2>K$. Therefore, the second line of (\ref{chehel}) approaches zero from below, which proves that the support of $X_2^*$ is bounded away from $+\infty$. As mentioned before, a similar argument holds when $x_2\to -\infty$.
This proves that $X_2^*$ has a bounded support.

\textbf{Remark 3.} We remark here that the order of showing the boundedness of the supports is important. First, for a given $F_{X_2}$ (not necessarily bounded), it is proved that $F_{X_1}^*$ is bounded. Then, for a given bounded $F_{X_1}$, it is shown that $F_{X_2}^*$ is also bounded. The order is reversed when $\lambda>1$, and it follows the same steps as in the case of $\lambda\leq 1$. Therefore, it is omitted.

{\color{black}We next prove the second claim in Proposition 3. We assume that $0<\lambda< 1$, and a bounded $F_{X_1}$ is given. We already know that for a given bounded $F_{X_1}$, $F_{X_2}^*$ has a bounded support denoted by $[A_1,A_2]$. Therefore,
\begin{align}
I^*_{F_{X_1}}=\sup_{\substack{F_{X_2}: \\ \mathds{E}[X_2^2]\leq P_2'}}I_{\lambda}(F_{X_1}F_{X_2})\nonumber\\
I^*_{F_{X_1}}=\sup_{\substack{F_{X_2}\in\mathscr{S}_2: \\ \mathds{E}[X_2^2]\leq P_2'}}I_{\lambda}(F_{X_1}F_{X_2}),
\end{align}
where $\mathscr{S}_2$ denotes the set of all probability distributions on the Borel sets of $[A_1,A_2]$. Let $p_0^*=p_Y(0;F_{X_1}F_{X_2}^*)$ denote the probability of the event $Y=0$, induced by $F_{X_2}^*$ and the given $F_{X_1}$. Also, let $P_2^*$ denote the second moment of $X_2$ under $F_{X_2}^*$.
The set
\begin{equation}
\mathscr{F}_2=\left\{F_{X_2}\in\mathscr{S}_2|\int_{A_1}^{A_2}p(0|x_2)dF_{X_2}(x_2)=p_0^*,\ \int_{A_1}^{A_2}x_2^2dF_{X_2}(x_2)=P_2^*\right\}
\end{equation}
is the intersection of $\mathscr{S}_2$ with two hyperplanes.\footnote{Note that $\mathscr{S}_2$ is convex and compact.}.
We can write
\begin{equation}\label{objective}
I^*_{F_{X_1}}=\sup_{F_{X_2}\in\mathscr{F}_2}I_\lambda(F_{X_1}F_{X_2}).
\end{equation}
Note that having $F_{X_2}\in\mathscr{F}_2$, the objective function in (\ref{objective}) becomes
\begin{align}
\underbrace{\lambda H(Y)}_{\mbox{constant}} +\underbrace{(1-\lambda)H(Y|X_2) - H(Y|X_1,X_2)}_{\mbox{linear in }F_{X_2}}\label{I1}.
\end{align}
Since the linear part is continuous and $\mathscr{F}_2$ is compact\footnote{The continuity of the linear part follows similarly to the continuity arguments in Appendix \ref{app1}. Note that this compactness is due to the closedness of the intersecting hyperplanes in $\mathscr{F}_2$, since a closed subset of a compact set is compact \cite{rudin}. The hyperplanes are closed due to continuity of $x_2^2$ and $p(0|x_2)$ (see (\ref{yekonim})).}, the objective function in (\ref{objective}) attains its maximum at an extreme point of $\mathscr{F}_2$, which, by Dubins' theorem, is a convex combination of at most three extreme points of $\mathscr{S}_2$. Since the extreme points of $\mathscr{S}_2$ are the CDFs having only one point of increase in $[A_1,A_2]$, we conclude that given any bounded $F_{X_1}$, $F_{X_2}^*$ has at most three mass points.

Now, assume that an arbitrary $F_{X_2}$ is given with at most three mass points denoted by $\{x_{2,i}\}_{i=1}^3$. It is already known that the support of $F_{X_1}^*$ is bounded, which is denoted by $[A_1',A_2']$. Let $\mathscr{S}_1$ denote the set of all probability distributions on the Borel sets of $[A_1',A_2']$.
The set
\begin{align}
\mathscr{F}_1=\bigg\{F_{X_1}\in\mathscr{S}_1\bigg |\int_{A_1'}^{A_2'}p(0|x_1,x_{2,j})dF_{X_1}(x_1)=p(0;F_{X_1}^*|x_{2,j}),\ j\in[1:3],\ \int_{A_1'}^{A_2'}x_1^2dF_{X_1}(x_1)=P_1'
\bigg\},
\end{align}
is the intersection of $\mathscr{S}_1$ with four hyperplanes\footnote{Note that here, since we know $\theta_1\neq 0$, the optimal input attains its maximum power of $P_1'$.}. In a similar way,
\begin{equation}\label{objective2}
I^*_{F_{X_2}}=\sup_{F_{X_1}\in\mathscr{F}_1}\left\{I_\lambda(F_{X_1}F_{X_2})\right\},
\end{equation}
and having $F_{X_1}\in\mathscr{F}_1$, the objective function in (\ref{objective2}) becomes
\begin{align}
I_\lambda &=\underbrace{\lambda H(Y) +(1-\lambda)\sum_{i=1}^{3}p_{X_2}(x_{2,i})H(Y|X_2=x_{2,i})}_{\mbox{constant}} - \underbrace{H(Y|X_1,X_2)}_{\mbox{linear in }F_{X_1}}\label{I2}
\end{align}
Therefore, given any $F_{X_2}$ with at most three points of increase, $F_{X_1}^*$ has at most five mass points.

When $\lambda = 1$, the second term on the RHS of (\ref{I2}) disappears, which means that $\mathscr{F}_1$ could be replaced by
\begin{equation*}
\left\{F_{X_1}\in\mathscr{S}_1|\int_{A_1'}^{A_2'}p(0|x_1)dF_{X_1}(x_1)=\tilde{p}_0^*,\ \int_{A_1'}^{A_2'}x_1^2dF_{X_1}(x_1)=P_1'\right\},
\end{equation*}
where $\tilde{p}_0^*=p_Y(0;F_{X_1}^*F_{X_2})$ is the probability of the event $Y=0$, which is induced by $F_{X_1}^*$ and the given $F_{X_2}$. Since the number of intersecting hyperplanes has been reduced to two, it is concluded that $F_{X_1}^*$ has at most three points of increase.}
\end{proof}

{\color{black}
\textbf{Remark 4.} Note that, the order of showing the discreteness of the support sets is also important. First, for a given bounded $F_{X_1}$ (not necessarily discrete), it is proved that $F_{X_2}^*$ is discrete with at most three mass points. Then, for a given discrete $F_{X_2}$ with at most three mass points, it is shown that $F_{X_1}^*$ is also discrete with at most five mass points when $\lambda<1$, and at most three mass points when $\lambda=1$. When $\lambda>1$, the order is reversed and it follows the same steps as in the case of $\lambda<1$. Therefore, it is omitted.}

\section{Conclusion}\label{con}
We have studied the capacity region of a two-transmitter Gaussian MAC under average input power constraints and one-bit ADC front end at the receiver. We have shown that an auxiliary random variable is necessary for characterizing the capacity region in general. We have derived an upper bound on the cardinality of this auxiliary variable, and proved that the distributions that achieve the boundary points of the capacity region are finite and discrete.
%Furthermore, time division with power control has been shown to achieve the sum capacity of this channel. Finally, we settled the conjecture of the sufficiency of $K$ mass points in a point to point AWGN channel with a $K$-bin quantizer at the receiver.
\appendices
\section{}\label{appp1}
{\color{black}Since $|\mathcal{U}|\leq 5$, we assume $\mathcal{U}=\{0,1,2,3,4\}$ without loss of generality, since what matters in the evaluation of the capacity region is the mass probability of the auxiliary random variable $U$, not its actual values.

%Let $\Omega$ be defined as
%\begin{equation}
%\Omega = \big\{F_{U,X_1,X_2}\big|\ U\in\mathcal{U},\ X_1-U-X_2,\ \mathds{E}[X_j^2]\leq P_j,\ j=1,2\big\},
%\end{equation}
%which is the set of all CDFs on the triplet $(U,X_1,X_2)$, where $U$ is drawn from $\mathcal{U}$, the Markove chain $X_1-U-X_2$ and the corresponding average power constraints hold.

In order to show the compactness of $\Omega$, we adopt a general form of the approach in \cite{Abou}.

First, we show that $\Omega$ is tight\footnote{A set of probability distributions $\Theta$ defined on $\mathds{R}^k$, i.e. the set of CDFs $F_{X_1,X_2,\ldots,X_k}$, is said to be tight, if for every $\epsilon>0$, there is a compact set $K_{\epsilon}\subset\mathds{R}^k$ such that \cite{Shiryaev}
\begin{equation*}
\mbox{Pr}\big\{(X_1,X_2,\ldots,X_k)\in \mathds{R}^k\backslash K_\epsilon\big\}<\epsilon,\ \forall F_{X_1,X_2,\ldots,X_k}\in\Theta.
\end{equation*} }. Choose $T_j$, $j=1,2$, such that $T_j>\sqrt{\frac{2P_j}{\epsilon}}$. Then, from Chebyshev's inequality,
\begin{equation}\label{cheb1}
\mbox{Pr}\big\{|X_j|>T_j\big\}\leq \frac{P_j}{T_j^2}<\frac{\epsilon}{2},\ j=1,2.
\end{equation}
Let $K_\epsilon=[0,4]\times[-T_1,T_1]\times[-T_2,T_2]\subset\mathds{R}^3$. It is obvious that $K_\epsilon$ is a closed and bounded subset of $\mathds{R}^3$, and therefore, compact. With this choice of $K_\epsilon$, we have
\begin{align}
\mbox{Pr}\big\{(U,X_1,X_2)\in \mathds{R}^3\backslash K_\epsilon\big\}&\leq\mbox{Pr}\{U\notin[0,4]\}+\mbox{Pr}\{X_1\notin[-T_1,T_1]\}+\mbox{Pr}\{X_2\notin[-T_2,T_2]\}\nonumber\\
&<0+\frac{\epsilon}{2}+\frac{\epsilon}{2}=\epsilon,\label{cheb2}
\end{align}
where (\ref{cheb2}) is due to (\ref{cheb1}). Hence, $\Omega$ is tight.

From Prokhorov's theorem \cite[p.318]{Shiryaev}, a set of probability distributions is tight if and only if it is relatively sequentially compact\footnote{A subset of topological space is relatively compact if its closure is compact.}. This means that for every sequence of CDFs $\{F_n\}$ in $\Omega$, there exists a subsequence $\{F_{n_k}\}$ that is weakly convergent\footnote{The weak convergence of $\{F_n\}$ to $F$ (also shown as $F_n(x)\stackrel{w}\to F(x)$) is equivalent to
\begin{equation}
\lim_{n\to\infty}\int_{\mathbb{R}}\psi(x)dF_n(x)=\int_{\mathbb{R}}\psi(x)dF(x),
\end{equation}
for all continuous and bounded functions $\psi(\cdot)$ on $\mathbb{R}$. Note that $F_n(x)\stackrel{w}\to F(x)$ if and only if $d_L(F_n,F)\to 0$.} to a CDF $F_0$, which is not necessarily in $\Omega$. If we can show that this $F_0$ is also an element of $\Omega$, then the proof is complete, since we have shown that $\Omega$ is sequentially compact, and therefore, compact\footnote{Compactness and sequentially compactness are equivalent in metric spaces. Note that $\Omega$ is a metric space with L\'{e}vy distance.}.

Assume a sequence of distributions $\{F_n(\cdot,\cdot,\cdot)\}$ in $\Omega$ that converges weakly to $F_0(\cdot,\cdot,\cdot)$. In order to show that this limiting distribution is also in $\Omega$, we need to show that both the average power constraints and the Markov chain ($X_1-U-X_2$) are preserved under $F_0$. The preservation of the second moment follows similarly to the argument in \cite[Appendix I]{Abou}. In other words, since $x^2$ is continuous and bounded below, from \cite[Theorem 4.4.4]{Chung}
\begin{align}
\int x_j^2d^3F_0(u,x_1,x_2)&\leq\liminf_{n\to\infty}\int x_j^2d^3F_n(u,x_1,x_2)\nonumber\\
&\leq P_j,\ j=1,2,
\end{align}
Therefore, the second moments are preserved under the limiting distribution $F_0$.

For the preservation of the Markov chain $X_1-U-X_2$, we need the following proposition.

\textbf{Proposition 4.} Assume a sequence of distributions $\{F_n(\cdot,\cdot)\}$ over the pair of random variables $(X,Y)$ that converges weakly to $F_0(\cdot,\cdot)$. Also, assume that $Y$ has a finite support, i.e., $\mathcal{Y}=\{1,2,\ldots,|\mathcal{Y}|\}$. Then, the sequence of conditional distributions (conditioned on $Y$) converges weakly to the limiting conditional distribution (conditioned on $Y$), i.e.,
\begin{equation}\label{cong}
F_n(\cdot|y)\stackrel{w}\to F_0(\cdot|y),\ \forall y\in\mathcal{Y}.
\end{equation}
\begin{proof}
The proof is by contradiction. If (\ref{cong}) is not true, then there exists $y'\in\mathcal{Y}$, such that $F_n(\cdot|y')\cancel{\stackrel{w}\to} F_0(\cdot|y')$. This means, from the definition of weak convergence, that there exists a bounded continuous function of $x$, denoted by $g_{y'}(x)$, such that
\begin{equation}
\int g_{y'}(x)dF_n(x|y')\cancel{\to}\int g_{y'}(x)dF_0(x|y').
\end{equation}
Let $f(x,y)$ be a bounded continuous function that satisfies
\begin{equation}
f(x,y)=\left\{\begin{array}{cc} 0 & y\in\mathcal{Y},\ y\neq y'\\ g_{y'}(x) & y=y' \end{array}\right..
\end{equation}
With this choice of $f(x,y)$, we have
\begin{equation}
\int f(x,y)d^2F_n(x,y)\cancel{\to}\int f(x,y)d^2F_0(x,y),
\end{equation}
which violates the assumption of the weak convergence of $F_n(\cdot,\cdot)$ to $F_0(\cdot,\cdot)$. Therefore, (\ref{cong}) holds.
\end{proof}
Since $\{F_n(\cdot,\cdot,\cdot)\}$ in $\Omega$ converges weakly to $F_0(\cdot,\cdot,\cdot)$ and $\mathcal{U}$ is finite, from Proposition 4, we have
\begin{equation}
F_n(\cdot,\cdot|u)\stackrel{w}\to F_0(\cdot,\cdot|u),\ \forall u\in\mathcal{U},
\end{equation}
where it is obvious that the arguments are $x_1$ and $x_2$. Since $F_n\in\Omega$, we have $F_n(x_1,x_2|u)=F_n(x_1|u)F_n(x_2|u)\ \forall u\in\mathcal{U}$. Also, since the convergence of the joint distribution implies the convergence of the marginals, we have \cite{billingsley}, \cite[Theorem 2.7]{Sagitov},
\begin{equation}\label{sakh}
F_0(x_1,x_2|u)=F_0(x_1|u)F_0(x_2|u)\ \forall u\in\mathcal{U},
\end{equation}
which states that under the limiting distribution $F_0$, the Markov chain $X_1-U-X_2$ is preserved.\footnote{Alternatively, this could be proved by the lower-semicontinuity of the mutual information as follows.
\begin{align}
I_{F_0}(X_1;X_2|U=u)&\leq\liminf_{n\to\infty}I_{F_n}(X_1;X_2|U=u)\\
&=0,\ \forall u\in\mathcal{U},
\end{align}
where $I_F$ denotes the mutual information under distribution $F$. The last equality is from the conditional independence of $X_1$ and $X_2$ given $U=u$ under $F_n$. Therefore, $I_{F_0}(X_1;X_2|U=u)=0,\ \forall u\in\mathcal{U}$, which is equivalent to (\ref{sakh}).} This completes the proof of the compactness of $\Omega$.}
\section{Proof of Proposition 2}\label{app1}
\subsection{Concavity}
When $0<\lambda\leq 1$, we have
\begin{equation}
I_\lambda(F_{X_1}F_{X_2})=\lambda H(Y)+(1-\lambda)H(Y|X_2)-H(Y|X_1,X_2).\label{se}
\end{equation}
For a given $F_{X_1}$, $H(Y)$ is a concave function of $F_{X_2}$, while $H(Y|X_2)$ and $H(Y|X_1,X_2)$ are linear in $F_{X_2}$. Therefore, $I_\lambda$ is a concave function of $F_{X_2}$. For a given $F_{X_2}$, $H(Y)$ and $H(Y|X_2)$ are concave functions of $F_{X_1}$, while $H(Y|X_1,X_2)$ is linear in $F_{X_1}$. Since $(1-\lambda)\geq 0$, $I_\lambda$ is a concave function of $F_{X_1}$. The same reasoning applies to the case $\lambda>1$.
\subsection{Continuity}
When $\lambda\leq 1$, the continuity of the three terms on the RHS of (\ref{se}) is investigated. Let $\{F_{X_2,n}\}$ be a sequence of distributions which is weakly convergent to $F_{X_2}$. For a given $F_{X_1}$, we have
\begin{align}
\lim_{x_2\to x_2^0}p(y;F_{X_1}|x_2)&=\lim_{x_2\to x_2^0}\int Q(x_1+x_2)dF_{X_1}(x_1)\nonumber\\
&=\int\lim_{x_2\to x_2^0}Q(x_1+x_2)dF_{X_1}(x_1)\label{yek}\\
&=p(y;F_{X_1}|x_2^0),\label{yekonim}
\end{align}
where (\ref{yek}) is due to the fact that the $Q$ function can be dominated by 1, which is an absolutely integrable function over $F_{X_1}$. Therefore, $p(y;F_{X_1}|x_2)$ is continuous in $x_2$, and combined with the weak convergence of $\{F_{X_2,n}\}$, we can write
\begin{align*}
\lim_{n\to\infty}p(y;F_{X_1}F_{X_2,n})&=\lim_{n\to\infty}\int p(y;F_{X_1}|x_2)dF_{X_2,n}(x_2)\nonumber\\&=\int p(y;F_{X_1}|x_2)dF_{X_2}(x_2)\nonumber\\
&= p(y;F_{X_1}F_{X_2}).
\end{align*}
This allows us to write
\begin{align*}
\lim_{n\to\infty}-\sum_{y=0}^{1}p(y;F_{X_1}F_{X_2,n})\log p(y;F_{X_1}F_{X_2,n})=-\sum_{y=0}^{1}p(y;F_{X_1}F_{X_2})\log p(y;F_{X_1}F_{X_2}),
\end{align*}
which proves the continuity of $H(Y)$ in $F_{X_2}$. $H(Y|X_2=x_2)$ is a bounded ($\in[0,1]$) continuous function of $x_2$, since it is a continuous function of $p(y;F_{X_1}|x_2)$, and the latter is continuous in $x_2$ (see (\ref{yekonim})). Therefore,
\begin{equation*}
\lim_{n\to\infty}\int H(Y|X_2=x_2)dF_{X_2,n}(x_2)=\int H(Y|X_2=x_2)dF_{X_2}(x_2),
\end{equation*}
which proves the continuity of $H(Y|X_2)$ in $F_{X_2}$. In a similar way, it can be verified that $\int H(Y|X_1=x_1,X_2=x_2)dF_{X_1}(x_1)$ is a bounded and continuous function of $x_2$ which guarantees the continuity of $H(Y|X_1,X_2)$ in $F_{X_2}$, since
\begin{equation}
H(Y|X_1,X_2)=\int\left(\int H(Y|X_1=x_1,X_2=x_2)dF_{X_1}(x_1)\right)dF_{X_2}(x_2)
\end{equation}
Therefore, for a given $F_{X_1}$, $I_\lambda$ is a continuous function of $F_{X_2}$. Exchanging the roles of $F_{X_1}$ and $F_{X_2}$ and also the case $\lambda>1$ can be addressed similarly, and are omitted for the sake of brevity.
\subsection{Weak Differentiability}
%$I_{\lambda}(F_{X_1}F_{X_2})$ ($0<\lambda\leq 1$) can be written as
%\begin{align}
%I_\lambda(F_{X_1}F_{X_2}) &=\int_{-\infty}^{+\infty}\int_{-\infty}^{+\infty}\sum_{y=0}^{1}p(y|x_1,x_2)\log\frac{p(y|x_1,x_2)}{[p(y;F_{X_1}F_{X_2})]^\lambda[p(y;F_{X_1}|x_2)]^{1-\lambda}}dF_{X_1}(x_1)dF_{X_2}(x_2)\nonumber\\
%& = \int_{-\infty}^{+\infty}\tilde{i}_\lambda(x_1;F_{X_1}|F_{X_2})dF_{X_1}(x_1)\label{mar1}\\
%& =\int_{-\infty}^{+\infty}i_\lambda(x_2;F_{X_2}|F_{X_1})dF_{X_2}(x_2),
%\end{align}
%%The notation $p(y;F_{X_1}F_{X_2})$ emphasizes the fact that the pmf of $Y$ has been induced by the product $F_{X_1}F_{X_2}$. Similarly, $p(y;F_{X_1}|x_2)$ shows that the conditional pmf of $Y$ conditioned on $x_2$ has been induced by $F_{X_1}$ (i.e., $p(y;F_{X_1}|x_2)=\int p(y|x_1,x_2)dF_{X_1}$).
%where we have defined
%\begin{equation}
%\tilde{i}_\lambda(x_1;F_{X_1}|F_{X_2})\triangleq\int_{-\infty}^{+\infty}\left(D\left(p(y|x_1,x_2)||p(y;F_{X_1}F_{X_2})\right)+(1-\lambda)\sum_{y=0}^{1}p(y|x_1,x_2)\log\frac{p(y;F_{X_1}F_{X_2})}{p(y;F_{X_1}|x_2)}\right)dF_{X_2}(x_2),\label{dden1}
%\end{equation}
%and
%\begin{equation}
%i_\lambda(x_2;F_{X_2}|F_{X_1})\triangleq\int_{-\infty}^{+\infty}D\left(p(y|x_1,x_2)||p(y;F_{X_1}F_{X_2})\right)dF_{X_1}(x_1)-(1-\lambda)D\left(p(y;F_{X_1}|x_2)||p(y;F_{X_1}F_{X_2})\right).\label{dden2}
%\end{equation}
%Here, (\ref{dden1}) can be considered as the density of $I_\lambda$ over $F_{X_1}$ when $F_{X_2}$ is given. (\ref{dden2}) can be interpreted in a similar way.

For a given $F_{X_1}$, the weak derivative of $I_\lambda$ at $F_{X_2}^0$ is given by
\begin{equation}
I'_\lambda(F_{X_1}F_{X_2})|_{F_{X_2}^0}=\lim_{\beta\to 0^+}\frac{I_\lambda(F_{X_1}((1-\beta)F_{X_2}^0+\beta F_{X_2}))-I_\lambda(F_{X_1}F_{X_2}^0)}{\beta},
\end{equation}
if the limit exists. It can be verified that
\begin{align*}
I'_\lambda(F_{X_1}F_{X_2})|_{F_{X_2}^0}&=\lim_{\beta\to 0^+}\frac{\int i_\lambda(x_2;(1-\beta)F_{X_2}^0+\beta F_{X_2}|F_{X_1})d((1-\beta)F_{X_2}^0(x_2)+\beta F_{X_2}(x_2))-\int i_\lambda(x_2;F_{X_2}^0|F_{X_1})dF_{X_2}^0(x_2)}{\beta}\nonumber\\
&=\int i_\lambda(x_2;F_{X_2}^0|F_{X_1})dF_{X_2}(x_2)-\int i_\lambda(x_2;F_{X_2}^0|F_{X_1})dF_{X_2}^0(x_2)\nonumber\\
&=\int i_\lambda(x_2;F_{X_2}^0|F_{X_1})dF_{X_2}(x_2)-I_\lambda(F_{X_1}F_{X_2}^0),
\end{align*}
where $i_\lambda$ has been defined in (\ref{dden2}). In a similar way, for a given $F_{X_2}$, the weak derivative of $I_\lambda$ at $F_{X_1}^0$ is
\begin{equation}\label{negah}
I'_\lambda(F_{X_1}F_{X_2})|_{F_{X_1}^0}=\int \tilde{i}_\lambda(x_1;F_{X_1}^0|F_{X_2})dF_{X_1}(x_1)-I_\lambda(F_{X_1}^0F_{X_2}),
\end{equation}
where $\tilde{i}_\lambda$ has been defined in (\ref{dden1}). The case $\lambda>1$ can be addressed similarly.
\section{Proof of Lemma 2}\label{app2}
We have
\begin{align}
\sup_{\substack{F_{X_1}: \\ \mathds{E}[X_1^2]\leq P_1'}}I_{\lambda}(F_{X_1}F_{X_2})&\leq\sup_{\substack{F_{X_1}F_{X_2}: \\ \mathds{E}[X_j^2]\leq P_j',\ j=1,2}}I_{\lambda}(F_{X_1}F_{X_2})\nonumber\\
&\leq \sup_{\substack{F_{X_1}F_{X_2}: \\ \mathds{E}[X_j^2]\leq P_j',\ j=1,2}}I(X_1,X_2;Y)\label{dfg}\\
&\leq \sup_{\substack{F_{X_1}F_{X_2}: \\ \mathds{E}[X_j^2]\leq P_j',\ j=1,2}}H(Y)-\inf_{\substack{F_{X_1}F_{X_2}: \\ \mathds{E}[X_j^2]\leq P_j',\ j=1,2}}H(Y|X_1,X_2)\nonumber\\
&= 1 - \inf_{\substack{F_{X_1}F_{X_2}: \\ \mathds{E}[X_j^2]\leq P_j',\ j=1,2}}\int\!\!\int H_b\left(Q(x_1+x_2)\right)dF_{X_1}(x_1)dF_{X_2}(x_2)\nonumber\\
&=1 - \inf_{\substack{F_{X_1}F_{X_2}: \\ \mathds{E}[X_j^2]\leq P_j',\ j=1,2}}\int\!\!\int H_b\left(Q\left(\sqrt{x_1^2}+\sqrt{x_2^2}\right)\right)dF_{X_1}(x_1)dF_{X_2}(x_2)\label{mrs}\\
&\leq 1 - \inf_{\substack{F_{X_1}F_{X_2}: \\ \mathds{E}[X_j^2]\leq P_j',\ j=1,2}}\int\!\!\int Q\left(\sqrt{x_1^2}+\sqrt{x_2^2}\right)dF_{X_1}(x_1)dF_{X_2}(x_2)\label{mrs.}\\
&=1-Q\left(\sqrt{P_1'}+\sqrt{P_2'}\right)\label{mrs2}\\
&<1,
\end{align}
where (\ref{dfg}) is from the non-negativity of mutual information and the assumption that $0<\lambda\leq 1$; (\ref{mrs}) is justified since the $Q$ function is monotonically decreasing and the sign of the inputs does not affect the average power constraints, $X_1$ and $X_2$ can be assumed non-negative (or alternatively non-positive) without loss of optimality; in (\ref{mrs.}), we use the fact that $Q\left(\sqrt{x_1^2}+\sqrt{x_2^2}\right)\leq\frac{1}{2}$, and for $t\in[0,\frac{1}{2}]$, $H_b(t)\geq t$; (\ref{mrs2}) is based on the convexity and monotonicity of the function $Q(\sqrt{u}+\sqrt{v})$ in $(u,v)$, which is shown in Appendix \ref{app4}. Therefore, the LHS of (\ref{khalaf}) is strictly less than 1.

Since $X_2$ has a finite second moment ($\mathds{E}[X_2^2]\leq P_2'$), from Chebyshev's inequality, we have
\begin{equation}\label{mori}
P(|X_2|\geq M)\leq \frac{P_2'}{M^2},\ \ \forall M>0.
\end{equation}
Fix $M>0$ and consider $X_1\sim F_{X_1}(x_1)=\frac{1}{2}[s(x_1+2M)+s(x_1-2M)]$. By this choice of $F_{X_1}$, we get
\begin{align}
I_\lambda(F_{X_1}F_{X_2})&=I(X_1;Y|X_2)+\lambda I(X_2;Y)\nonumber\\
&\geq I(X_1;Y|X_2)\label{noghte}\\
&=\int_{-\infty}^{+\infty}I(X_1;Y|X_2=x_2)dF_{X_2}(x_2)\nonumber\\
&\geq \int_{-M}^{+M}I(X_1;Y|X_2=x_2)dF_{X_2}(x_2)
\end{align}
\begin{align}
&\geq \inf_{F_{X_2}}\int_{-M}^{+M}H(Y|X_2=x_2)dF_{X_2}(x_2)-\sup_{F_{X_2}}\int_{-M}^{+M}H(Y|X_1,X_2=x_2)dF_{X_2}(x_2)\nonumber\\
&\geq\left(1-\frac{P_2'}{M^2}\right)H_b\left(\frac{1}{2}-\frac{1}{2}(Q(3M)+Q(M))\right)-H_b\left(Q(2M)\right)\label{sishish},
\end{align}
where (\ref{sishish}) is due to (\ref{mori}) and the fact that $H(Y|X_2=x_2)=H_b(\frac{1}{2}Q(2M+x_2)+\frac{1}{2}Q(-2M+x_2))$ is minimized over $[-M,M]$ at $x_2=M$ (or, alternatively at $x_2=-M$), and $H(Y|X_1,X_2=x_2)=\frac{1}{2}H_b(Q(2M+x_2))+\frac{1}{2}H_b(Q(-2M+x_2))$ is maximized at $x_2=0$. (\ref{sishish}) shows that $I_\lambda$ ($\leq 1$) can become arbitrarily close to 1 given that $M$ is large enough. Hence, its supremum over all distributions $F_{X_1}$ is 1. This means that (\ref{khalaf}) cannot hold, and $\theta_1\neq 0$.
\section{Justification of (\ref{nec}), (\ref{nec2}) and (\ref{nec3})}\label{just}
{\color{black}Let $X$ be a vector space, and $Z$ be a real-valued function defined on a convex domain $D\subset X$. Suppose that $x^*$ maximizes $Z$ on $D$, and that $Z$ is Gateaux differentiable (weakly differentiable) at $x^*$. Then, from \cite[Th.2, p.178]{Luenberger},
\begin{equation}\label{lue}
Z'(x)|_{x^*}\leq 0,
\end{equation}
where $Z'(x)|_{x^*}$ is the weak derivative of $Z$ at $x^*$.

From (\ref{negah}), we have the weak derivative of $I_{\lambda}$ at $F^*_{X_1}$ as
\begin{equation}
I'_{\lambda}(F_{X_1}F_{X_2})|_{F^*_{X_1}}=\int \tilde{i}_\lambda(x_1;F_{X_1}^*|F_{X_2})dF_{X_1}(x_1)-I_\lambda(F_{X_1}^*F_{X_2}).
\end{equation}
Now, the derivation of (\ref{nec}) is immediate by inspecting that the weak derivative of the objective of (\ref{yekom}) at $F^*_{X_1}$ is given by
\begin{align}
I'_{\lambda}(F_{X_1}F_{X_2})|_{F^*_{X_1}}-\theta_1\left(\int x_1^2dF_{X_1}(x_1)-\int x_1^2dF^*_{X_1}(x_1)\right)&=\int \tilde{i}_\lambda(x_1;F_{X_1}^*|F_{X_2})dF_{X_1}(x_1)-I_\lambda(F_{X_1}^*F_{X_2})\nonumber\\
& -\theta_1\left(\int x_1^2dF_{X_1}(x_1)-\int x_1^2dF^*_{X_1}(x_1)\right).\label{has}
\end{align}
Letting (\ref{has}) be lower than or equal to zero (as in (\ref{lue})) results in (\ref{nec}).

The equivalence of (\ref{nec}) to (\ref{nec2}) and (\ref{nec3}) follows similarly to the proof of Corollary 1 in \cite[p.210]{Smith}.}
\section{Proof of Lemma 3}\label{app3}
(\ref{Le1}) is obtained as follows.
\begin{align}
\bigg|D\big(p(y|x_1,x_2)||p(y;F_{X_1}F_{X_2})\big)\bigg|&=\left|\sum_{y=0}^{1}p(y|x_1,x_2)\log\frac{p(y|x_1,x_2)}{p(y;F_{X_1}F_{X_2})}\right|\nonumber\\&\leq\bigg|H(Y|X_1=x_1,X_2=x_2)\bigg|+\left|\sum_{y=0}^{1}p(y|x_1,x_2)\log p(y;F_{X_1}F_{X_2})\right|\nonumber\\
&\leq 1 +\left|\sum_{y=0}^{1}\log p(y;F_{X_1}F_{X_2})\right|\label{uv}\\
&=1 -\sum_{y=0}^{1}\log p(y;F_{X_1}F_{X_2})\nonumber\\
&\leq 1- 2\min\bigg\{\log p_Y(0;F_{X_1}F_{X_2}),\log p_Y(1;F_{X_1}F_{X_2})\bigg\}\nonumber\\
&\leq 1- 2\log Q(\sqrt{P_1'}+\sqrt{P_2'})\label{jh}\\&<\infty\nonumber,
\end{align}
where (\ref{uv}) is due to the fact that the binary entropy function is upper bounded by 1. (\ref{jh}) is justified as follows.
\begin{align}
\min\bigg\{ p_Y(0;F_{X_1}F_{X_2}), p_Y(1;F_{X_1}F_{X_2})\bigg\}&\geq\inf_{\substack{F_{X_1}F_{X_2}: \\ \mathds{E}[X_j^2]\leq P_j'}}\min\bigg\{ p_Y(0;F_{X_1}F_{X_2}), p_Y(1;F_{X_1}F_{X_2})\bigg\}\nonumber\\
&=\inf_{\substack{F_{X_1}F_{X_2}: \\ \mathds{E}[X_j^2]\leq P_j'}} p_Y(0;F_{X_1}F_{X_2})\nonumber\\
&=\inf_{\substack{F_{X_1}F_{X_2}: \\ \mathds{E}[X_j^2]\leq P_j'}}\int\!\!\int Q(x_1+x_2)dF_{X_1}(x_1)dF_{X_2}(x_2)\nonumber\\
&=\inf_{\substack{F_{X_1}F_{X_2}: \\ \mathds{E}[X_j^2]\leq P_j'}}\int\!\!\int Q\left(\sqrt{x_1^2}+\sqrt{x_2^2}\right)dF_{X_1}(x_1)dF_{X_2}(x_2)\label{uu}\\
&\geq Q\left(\sqrt{P_1'}+\sqrt{P_2'}\right)\label{vv},
\end{align}
where (\ref{vv}) is based on the convexity and monotonicity of the function $Q(\sqrt{u}+\sqrt{v})$, which is shown in appendix \ref{app4}.

(\ref{Le2}) is obtained as follows.
\begin{align}
p(y;F_{X_1}|x_2)&\geq \min\bigg\{p(0;F_{X_1}|x_2),p(1;F_{X_1}|x_2)\bigg\}\nonumber\\
&\geq \int Q\left(|x_1|+|x_2|\right)dF_{X_1}(x_1)\nonumber\\
&=\int Q\left(\sqrt{x_1^2}+|x_2|\right)dF_{X_1}(x_1)\nonumber\\
&\geq Q\left(\sqrt{P_1'}+|x_2|\right)\label{akhari},
\end{align}
where (\ref{akhari}) is due to convexity of $Q(\alpha+\sqrt{x})$ in $x$ for $\alpha\geq 0$.

{\color{black}(\ref{Le3}) is obtained as follows.
\begin{align}
\left|\sum_{y=0}^{1}p(y|x_1,x_2)\log\frac{p(y;F_{X_1}F_{X_2})}{p(y;F_{X_1}|x_2)}\right|&\leq -\sum_{y=0}^{1}p(y|x_1,x_2)\log p(y;F_{X_1}|x_2)-\sum_{y=0}^{1}p(y|x_1,x_2)\log p(y;F_{X_1}F_{X_2})\nonumber\\
&\leq -\sum_{y=0}^{1}\log p(y;F_{X_1}|x_2)-\sum_{y=0}^{1}\log p(y;F_{X_1}F_{X_2})\label{halat}\\
&\leq -2\log Q\left(\sqrt{P_1'}+\sqrt{P_2'}\right)-2\log Q\left(\sqrt{P_1'}+|x_2|\right)\label{chera},
\end{align}
where (\ref{halat}) is from $p(y|x_1,x_2)\leq 1$; and (\ref{chera}) is
from (\ref{akhari}) and (\ref{vv}).

Note that, (\ref{chera}) is integrable with respect to $F_{X_2}$ due to the concavity of $-\log Q(\alpha+\sqrt{x})$ in $x$ for $\alpha\geq 0$ as shown in Appendix \ref{app4}. In other words,
\begin{align}
\int_{-\infty}^{+\infty}\left(-2\log Q\left(\sqrt{P_1'}+\sqrt{P_2'}\right)-2\log Q\left(\sqrt{P_1'}+|x_2|\right)\right)dF_{X_2}(x_2)&<-4\log Q\left(\sqrt{P_1'}+\sqrt{P_2'}\right)\\ &<+\infty.\label{xcxc}
\end{align}}
\section{Two convex functions}\label{app4}
%Let $f(x)=Q(a+\sqrt{x})$ for $x,a\geq 0$. We have,
%\begin{equation*}
%f'(x)=-\frac{e^{-\frac{(a+\sqrt{x})^2}{2}}}{2\sqrt{2\pi x}},
%\end{equation*}
%and
%\begin{equation*}
%f''(x)=\frac{\sqrt{x}+\frac{1}{\sqrt{x}}+a}{4\sqrt{2\pi}x}e^{-\frac{(a+\sqrt{x})^2}{2}}>0
%\end{equation*}
%which proves the (strict) convexity of $f(x)$.

Let $f(x)=\log Q(a+\sqrt{x})$ for $x,a\geq 0$. We have,
\begin{equation*}
f'(x)=-\frac{e^{-\frac{(a+\sqrt{x})^2}{2}}}{2\sqrt{2\pi x}Q(a+\sqrt{x})},
\end{equation*}
and
\begin{equation}\label{shast}
f''(x)=\frac{e^{-\frac{(a+\sqrt{x})^2}{2}}}{4x\sqrt{2\pi}Q^2(a+\sqrt{x})}\left((a+\sqrt{x}+\frac{1}{\sqrt{x}})Q(a+\sqrt{x})-\phi(a+\sqrt{x})\right),
\end{equation}
where $\phi(x)=\frac{1}{\sqrt{2\pi}}e^{-\frac{x^2}{2}}$. Note that
\begin{align}
(1+at+t^2)Q(a+t)+a\phi(a+t)&>\left(1+(a+t)^2\right)Q(a+t)\label{pb1}\\
&>(a+t)\phi(a+t)\label{pb2},\ \ \forall a,t>0,
\end{align}
where (\ref{pb1}) and (\ref{pb2}) are, respectively, due to $\phi(x)>xQ(x)$ and $(1+x^2)Q(x)>x\phi(x)$ ($x>0$). Therefore,
\begin{equation*}
(a+\sqrt{x}+\frac{1}{\sqrt{x}})Q(a+\sqrt{x})>\phi(a+\sqrt{x}),
\end{equation*}
which makes the second derivative in (\ref{shast}) positive, and proves the (strict) convexity of $f(x)$.

Let $f(u,v)=Q(\sqrt{u}+\sqrt{v})$ for $u,v\geq 0$. By simple differentiation, the Hessian matrix of $f$ is
\begin{equation}\label{GHIYAS}
\mathbf{H}=\frac{e^{-\frac{(\sqrt{u}+\sqrt{v})^2}{2}}}{\sqrt{2\pi}}\left[\begin{array}{cc} \frac{1}{2u\sqrt{u}}+\frac{\sqrt{u}+\sqrt{v}}{4u} & \frac{\sqrt{u}+\sqrt{v}}{4\sqrt{u}\sqrt{v}}\\ \frac{\sqrt{u}+\sqrt{v}}{4\sqrt{u}\sqrt{v}} & \frac{1}{2v\sqrt{v}}+\frac{\sqrt{u}+\sqrt{v}}{4v} \end{array}\right].
\end{equation}
It can be verified that $\mbox{det}(\mathbf{H})>0$ and $\mbox{trace}(\mathbf{H})>0$. Therefore, both eigenvalues of $\mathbf{H}$ are positive, which makes the matrix positive definite. Hence, $Q(\sqrt{u}+\sqrt{v})$ is (strictly) convex in $(u,v)$.
\section{Proof of lemma 4}\label{app5}
Let $A\triangleq \max\{A_1,A_2\}$.
\begin{figure}[t]
  \centering
  \includegraphics[width=12cm]{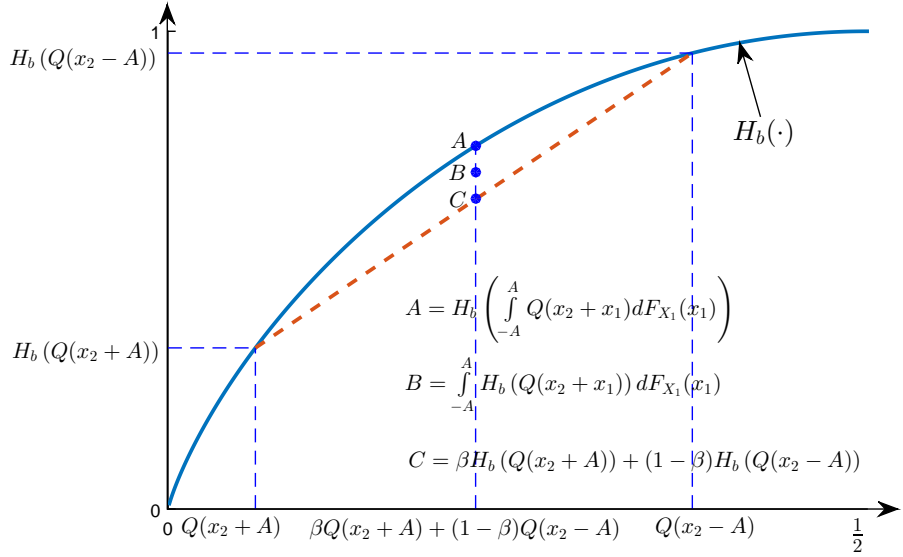}\\
  \caption{The figure depicting (\ref{haf}) and (\ref{eyey}). Note that in the statement of Lemma 4, $x_2\to+\infty$. Hence, we have assumed $x_2>A$ in the figure. }\label{fig3}
\end{figure}
It is obvious that
\begin{equation}
Q(x_2+A)\leq\int_{-A}^{A}Q(x_1+x_2)dF_{X_1}(x_1)\leq Q(x_2-A).
\end{equation}
Therefore, we can write
\begin{equation}\label{somebeta}
\int_{-A}^{A}Q(x_1+x_2)dF_{X_1}(x_1)= \beta Q(x_2+A)+(1-\beta)Q(x_2-A),
\end{equation}
for some $\beta\in[0,1].$ Note that $\beta$ is a function of $x_2$. Also, due to concavity of $H_b(\cdot)$, we have
\begin{align}\label{haf}
H_b\bigg(\int_{-A}^{A}Q(x_1+x_2)dF_{X_1}(x_1)\bigg)&\geq\int_{-A}^{A}H_b(Q(x_1+x_2))dF_{X_1}(x_1).
\end{align}
From the fact that
\begin{equation}\label{fac}
H_b(x)\geq \frac{H_b(p)-H_b(a)}{p-a}(x-a)+H_b(a),\ \forall x\in[a,p],\ \forall a,p\in[0,1] (a<p),
\end{equation}
we can also write
\begin{align}
\int_{-A}^{A}H_b(Q(x_1+x_2))dF_{X_1}(x_1)&\geq\frac{H_b(Q(x_2-A))\!-\!H_b(Q(x_2+A))}{Q(x_2-A)\!-\!Q(x_2+A)}\left(\int_{-A}^{A}Q(x_1+x_2)dF_{X_1}(x_1)\!-\!Q(x_2+A)\right)\nonumber\\
&\ \ \ +\!H_b(Q(x_2+A))\nonumber\\
&=\beta H_b(Q(x_2+A))+(1-\beta)H_b(Q(x_2-A))\label{eyey},
\end{align}
where (\ref{somebeta}) and (\ref{fac})  have been used in (\ref{eyey}). (\ref{haf}) and (\ref{eyey}) are depicted in Figure \ref{fig3}.

From (\ref{somebeta}) and (\ref{eyey}), we have
\begin{equation}\label{payeen}
\frac{\beta H_b(Q(x_2+A))+(1-\beta)H_b(Q(x_2-A))}{H_b\bigg(\beta Q(x_2+A)+(1-\beta)Q(x_2-A)\bigg)}\leq\frac{\int_{-A}^{A}H_b(Q(x_1+x_2))dF_{X_1}(x_1)}{H_b\bigg(\int_{-A}^{A} Q(x_1+x_2)dF_{X_1}(x_1)\bigg)}\leq 1.
\end{equation}
Let
\begin{equation}\label{defbeta}
\beta^*\triangleq\argmin_{\beta}\frac{\beta H_b(Q(x_2+A))+(1-\beta)H_b(Q(x_2-A))}{H_b\bigg(\beta Q(x_2+A)+(1-\beta)Q(x_2-A)\bigg)}.
\end{equation}
This minimizer satisfies the following equality
\begin{equation}\label{hala}
\frac{d}{d\beta}\left(\frac{\beta H_b(Q(x_2+A))+(1-\beta)H_b(Q(x_2-A))}{H_b\bigg(\beta Q(x_2+A)+(1-\beta)Q(x_2-A)\bigg)}\right)\bigg |_{\beta=\beta^*}=0.
\end{equation}
Therefore, we can write
\begin{align}
\frac{\beta H_b(Q(x_2+A))+(1-\beta)H_b(Q(x_2-A))}{H_b\bigg(\beta Q(x_2+A)+(1-\beta)Q(x_2-A)\bigg)}&\geq\frac{\beta^* H_b(Q(x_2+A))+(1-\beta^*)H_b(Q(x_2-A))}{H_b\bigg(\beta^* Q(x_2+A)+(1-\beta^*)Q(x_2-A)\bigg)}\label{haf1}\\
&= \frac{\frac{H_b(Q(x_2-A))-H_b(Q(x_2+A))}{Q(x_2-A)-Q(x_2+A)}}{H_b'\bigg(\beta^* Q(x_2+A)+(1-\beta^*)Q(x_2-A)\bigg)}\label{haf2}\\
&\geq\frac{\frac{H_b(Q(x_2-A))-H_b(Q(x_2+A))}{Q(x_2-A)-Q(x_2+A)}}{H_b'( Q(x_2+A))},\label{lab}
\end{align}
where (\ref{haf1}) is from the definition in (\ref{defbeta}); (\ref{haf2}) is from the expansion of (\ref{hala}), and $H_b'(t)=\log(\frac{1-t}{t})$ is the derivative of the binary entropy function; (\ref{lab}) is due to the fact that $H_b'(t)$ is a decreasing function.

Applying L'hospital's rule multiple times, we obtain
\begin{align}
\lim_{x_2\to+\infty}\frac{\frac{H_b(Q(x_2-A))-H_b(Q(x_2+A))}{Q(x_2-A)-Q(x_2+A)}}{H_b'( Q(x_2+A))}&=\lim_{x_2\to+\infty}\frac{H_b(Q(x_2-A))\bigg(1-\frac{H_b(Q(x_2+A))}{H_b(Q(x_2-A))}\bigg)}{Q(x_2-A)\bigg(1-\frac{Q(x_2+A)}{Q(x_2-A)}\bigg)\log(\frac{1- Q(x_2+A)}{Q(x_2+A)})}\nonumber\\&=\lim_{x_2\to+\infty}-\frac{H_b(Q(x_2-A))}{Q(x_2-A)\log( Q(x_2+A))}
\nonumber\\
&=\lim_{x_2\to+\infty}\frac{e^{-\frac{(x_2-A)^2}{2}}\log (Q(x_2-A))}{e^{-\frac{(x_2-A)^2}{2}}\log (Q(x_2+A))+\frac{Q(x_2-A)}{Q(x_2+A)}e^{-\frac{(x_2+A)^2}{2}}}\nonumber
\end{align}
\begin{align}
&=\lim_{x_2\to+\infty}\frac{\log(Q(x_2-A))}{\log(Q(x_2+A))+1}\nonumber\\
&\color{black}=\lim_{x_2\to+\infty}\frac{Q(x_2+A)e^{Ax_2}}{Q(x_2-A)e^{-Ax_2}}\nonumber\\
&=1\label{neshoon}
\end{align}

%where (\ref{had}) is due to the fact that for $\alpha>\beta$, we have
%\begin{equation*}
%\lim_{x_2\to+\infty}\frac{Q(\alpha+x_2)}{Q(\beta+x_2)}=\lim_{x_2\to+\infty}\frac{H_b(Q(\alpha+x_2))}{H_b(Q(\beta+x_2))}=0,
%\end{equation*}
%which is proved by the application of L'Hospital's rule.
From (\ref{payeen}), (\ref{lab}) and (\ref{neshoon}), (\ref{trr}) is proved. Note that the boundedness of $X_1$ is crucial in the proof. In other words, the fact that $Q(x_2-A)\to 0$ as $x_2\to+\infty$ is the very result of $A<+\infty$.

%Finally,
%\begin{equation}
%\lim_{x\to 0}\frac{H_b(x)}{cx}=\lim_{x\to 0}\frac{-\log x}{c}=+\infty,\ c>0.
%\end{equation}

\bibliography{REFERENCE}
\bibliographystyle{IEEEtran}
\end{document}